\documentclass[12pt,english]{article}
\usepackage[T1]{fontenc}
\usepackage[latin9]{inputenc}
\usepackage[letterpaper]{geometry}
\usepackage{color}
\usepackage{units}
\usepackage{amsmath}
\usepackage{graphicx}
\usepackage{amssymb}
\usepackage{esint}
\usepackage[authoryear]{natbib}

\makeatletter

\providecommand{\tabularnewline}{\\}

\def\argmin{\mathop{\mathrm{argmin}}}

\makeatother

\usepackage{babel}

\begin{document}

\title{\textbf{\textcolor{black}{MSE-based analysis of optimal tuning functions
predicts phenomena observed in sensory neurons}}}

\date{Steve Yaeli$^{1,2}$ and Ron Meir$^{1,2}$\\
{\normalsize 1 Department of Electrical Engineering}\\
{\normalsize 2 Laboratory for Network Biology Research}\\
{\normalsize Technion, Haifa}\\
{\normalsize Israel}\\
{\color{white}10.2.2010\color{black}}\\
{\textsc{Submitted to Frontiers in Computational Neuroscience}}\\
}

\maketitle
\textbf{}%
\begin{minipage}[t]{1\columnwidth}%
\textbf{Correspondence:} Ron Meir, Department of Electrical Engineering,
Technion,\\ Haifa 32000, Israel, rmeir@ee.technion.ac.il\\

\end{minipage}

\bigskip{}

\begin{abstract}
Biological systems display impressive capabilities in effectively
responding to environmental signals in real time. There is increasing
evidence that organisms may indeed be employing near optimal Bayesian
calculations in their decision-making. An intriguing question relates
to the properties of optimal encoding methods, namely determining
the properties of neural populations in sensory layers that optimize
performance, subject to physiological constraints. Within an ecological
theory of neural encoding/decoding, we show that optimal Bayesian
performance requires neural adaptation which reflects environmental
changes. Specifically, we predict that neuronal tuning functions possess
an optimal width, which increases with prior uncertainty and environmental
noise, and decreases with the decoding time window. Furthermore, even
for static stimuli, we demonstrate that dynamic sensory tuning functions,
acting at relatively short time scales, lead to improved performance.
Interestingly, the narrowing of tuning functions as a function of
time was recently observed in several biological systems. Such results
set the stage for a functional theory which may explain the high reliability
of sensory systems, and the utility of neuronal adaptation occurring
at multiple time scales.
\end{abstract}
\begin{minipage}[t]{1\columnwidth}%
\textbf{Keywords: }Tuning functions, neural encoding, population coding,
Bayesian decoding, optimal width, Fisher information. %
\end{minipage}

\section{\label{sec:Introduction}Introduction}

Behaving organisms often exhibit optimal, or nearly-optimal performance
(e.g., \citet{jacobs1999oit,ernst2002hiv,jacobs2009ror}), despite
physiological constraints and noise which is inherent to both the
environment and the neural activity. Experimental work performed over
the past few years has suggested that one means by which such effective
behavior is achieved is through adaptation to the environment occurring
across multiple time scal\textcolor{black}{es (e.g. milliseconds -
\citet{wang2005sfa}, seconds - \citealp{dean2008rna} and minutes
- \citet{pettet1992dcr}). }A major challenge pertaining to such results
relates to the construction of a coherent theoretical framework within
which the efficacy of such adaptation processes can be motivated and
assessed. We believe that a carefully articulated reverse engineering
perspective (see \citet{marom2009opp}), based on the utilization
of well founded engineering principles, can yield significant insight
about these phenomena. More specifically, such approaches can explain
how the environmental state may be effectively estimated from the
neural spike trains and from prior knowledge through a process of
neural decoding. Within this context, many psychophysical and neurophysiological
results can be explained by Bayesian models, suggesting that organisms
may indeed be employing Bayesian calculations in their decision-making
(\citet{knill1996pbi,knill2004tbb,rao2004bcr,ma2006bip}). Optimal
real-time decoding methods have been proposed over the past decade
for static (\citet{zemel1998pip,deneve2001ecc,pouget2002cpn,pouget2003cip,ma2006bip,averbeck2006ncp,beck2008ppc})
and dynamic (\citet{twum2001tep,eden2004dan,beck2007ein,huys2007fpc,pitkow2007ncv,deneve2008bsn,bobrowski2009bfi}\textcolor{black}{)}
environments. While much of this work has been performed in the context
of sensory systems, it has also been widely used in the context of
higher brain regi\textcolor{black}{ons (e.g., hippocampus - \citet{barbieri2004dai,eden2004dan}),
and may therefore be relevant across multiple} physiological levels.\\

Given the well developed theory of Bayesian decoding, an intriguing
follow-up question relates to the properties of optimal encoding methods,
namely determining the properties of a neural population in the sensory
layer that optimizes performance, subject to physiological constraints.
Sensory neurons are often characterized by their tuning functions
(sometimes referred to as ``tuning curves'' - e.g. \citet{anderson2000tcn,brenner2000arm,dragoi2000aip,harper2004onp,korte1993ast}),
which quantify the relationship between an external stimulus and the
evoked activities of each neuron, typically measured by the probability,
or frequency, of emitting spikes. From an ecological point of view
it is speculated that optimal tuning functions are not universal,
but rather adapt themselves to the specific context and to the statistical
nature of the environment. The neurophysiological literature offers
much experimental evidence for neurons in many areas, which respond
to changes in the statistical attributes of stimuli by modulating
their response properties (e.g. \citet{pettet1992dcr,brenner2000arm,dragoi2000aip,dean2005npc,hosoya2005dpc}).\\

Theoretical studies of optimal tuning functions hinge on the notion
of optimality, which requires the definition of an appropriate cost
function. Arguably, a natural cost function is some measure of distance
between the true and the estimated environmental state. In this context
a reasonable choice is the mean squared error (MSE), or the average
Euclidean distance between the two signals. While other distance measures
can be envisaged, and should indeed be studied in appropriate contexts,
there are many advantages to using this specific measure, not least
of which is its relative analytic tractability. The optimality criterion
then becomes minimal MSE (MMSE), where it is assumed that the spike
trains generated by the sensory cells are later processed by an optimal
decoder. Unfortunately, even in this relatively simple setting, analytical
examination of the dependence of the MMSE on the tuning functions
is infeasible in the general case, and many researchers draw conclusions
about optimality of tuning functions from lower bounds on the MSE,
especially those which make use of Fisher information (\citet{seung1993smr,pouget1999nvw,zhang1999nts,toyoizumi2002fis,johnson2004osc,series2004tcs,harper2004onp,lansky2005ose,brown2006ont}).
Unfortunately, in many cases reaching conclusions based on, often
loose, bounds can lead to very misleading results, which stand in
distinct opposition to predictions arising from analyzing the MMSE
itself (see Section \ref{sec:Discussion-1}). Along different lines,
other researchers consider information theoretic quantities (\citet{brunel1998mif,panzeri1999odr,mcdonnell2008mis,nikitin2009npc,geisler2009ose}),
attempting to determine conditions under which the maximal amount
of information is conveyed, subject to various constraints. However,
a direct relation (as opposed to bounds) between such quantities and
the MSE has only been established in very specific situations (\citet{duncan1970otc,guo2005mim})
and does not hold in general. A notable exception is the work of \citet{bethge2002ost}
and \citet{bethge2003onr}, who employed Monte Carlo simulations in
order to assess the MMSE directly in specific scenarios; see also
\citet{bobrowski2009bfi}, who computed the MMSE analytically to find
optimal tuning functions in a concrete setting.\\

In this paper we directly address the issue of MMSE-optimal tuning
functions, using well-justified approximations which lead to explicit
analytic results. We examine various scenarios, including some that
were not previously treated, and make novel predictions that can be
tested experimentally. We begin in Section \ref{sub:Fisher-optimal-width}
by analyzing optimality in terms of Fisher information and demonstrate
why drawing qualitative and quantitative conclusions based on bounds
can be misleading. In fact, bound-based predictions can sometimes
be diametrically opposed to the predictions based on the true error.
This notion is of great importance due to the very prevalent use of
approximations and bounds on the true error. We then move in Section
\ref{sub:MMSE-based-optimal-width} to discuss the implications of
directly minimizing the MMSE, focusing on the effects of noise and
multimodality. The advantages of dynamic real-time modification of
tuning function properties are analyzed in Section \ref{sub:Dynamic-optimal-width},
and concrete experimental predictions are made. In fact, some of these
predictions have already been observed in existing experimental data.
Specifically, we predict that neuronal tuning functions possess an
optimal width, which increases with prior uncertainty and environmental
noise, and decreases with the decoding time window. The results of
the paper are summarized and discussed in Section \ref{sec:Discussion-1},
and the mathematical details appear in Section \ref{sec:Methods}.
\\

\section{Results\label{sec:Results}}

We investigate the problem of neural encoding and decoding in a static
environment. More formally, consider an environment described by a
random variable $X$, taking values in some space ${\cal X}$, and
characterized by a probability density function $p(\cdot)$ (in order
to simplify notation, we refrain from indexing probability distributions
with the corresponding random variable, e.g., $p_{X}(\cdot)$, as
the random variable will be clear from the context ). In general,
$X$ may represent a stochastic dynamic process (e.g., \citet{bobrowski2009bfi})
but we limit ourselves in this study to the static case. In typical
cases $X$ may be a random vector, e.g., the spatial location of an
object, the frequency and intensity of a sound and so on. More generally,
in the case of neurons in associative cortical regions, $X$ can represent
more abstract quantities. Suppose that $X$ is sensed by a population
of $M$ sensory neurons which emit spike trains corresponding to the
counting processes $\mathbf{N}_{t}=\{N_{t}^{1},\ldots,N_{t}^{M}\}$,
where $N_{t}^{m}$ represents the number of spikes emitted by cell
$m$ up to time $t$. Denote by $\lambda_{m}(\cdot)$ the tuning function
of the \textit{m}-th sensory cell. We further assume that, given the
input, these spike trains are conditionally independent Poisson processes,
namely \[
N_{t}^{m}|X\sim\mathrm{Pois}\left(\lambda_{m}\left(X\right)t\right)\qquad(m=1,\ldots,M),\]
implying that $P(N_{t}^{m}=n|X)=e^{-\lambda_{m}(X)t}(\lambda_{m}(X)t)^{n}/n!$
(in a more general case of dynamic tuning functions $\{\lambda_{m}(t,\cdot)\}$,
with which we deal later, the parameter of the Poisson distribution
is $\int_{0}^{t}\lambda_{m}(s,X)ds$ - see \citet{bobrowski2009bfi}).
Following this encoding process, the goal of an optimal decoder is
to find the best reconstruction of the input $X$, based on observations
of the $M$ spike trains $\mathbf{N}_{t}$. In this paper we focus
on the converse problem, namely the selection of tuning functions
for encoding, that facilitate optimal decoding of $X$. \\

We formulate our results within a Bayesian setting, where it is assumed
that the input signal $X$ is drawn from some prior probability density
function (pdf) $p(\cdot)$, and the task is to estimate the true state
$X$, based on the prior distribution and the spike trains $\mathbf{N}_{t}$
observed up to time $t$. For any estimator $\hat{X},$ we consider
the mean squared error $\mathrm{MSE}\hat{(X)}=\mathbb{E}[(X-\hat{X})^{2}]$,
where the expectation is taken over both the observations $\mathbf{N}_{t}$
and the state $X.$ It is well known that $\hat{X}^{\text{opt}}$,
the estimator minimizing the MSE, is given by the conditional mean
$\mathbb{E}[X|\mathbf{N}_{t}]$ (e.g., \citet{vantrees1968dea}),
which depends on the parameters defining the tuning functions $\{\lambda_{m}(x)\}$.\\

As a specific example, assuming that the tuning functions are Gaussians
with centers $c_{1},\ldots,c_{M}$ and widths $\alpha_{1},\ldots,\alpha_{M}$,
the estimator $\hat{X}$, and the corresponding minimal MSE, referred
to as the MMSE, depend on $\boldsymbol{\theta}=(c_{1},\ldots,c_{M},\alpha_{1},\ldots,\alpha_{M})$.
The optimal values of the parameters are then given by a further minimization
process, \[
\boldsymbol{\boldsymbol{\theta}^{\mathrm{opt}}=\argmin_{\theta}}\left\{ \mathrm{MMSE}(\boldsymbol{\theta})\right\} .\]
In other words, $\boldsymbol{\theta}^{\mathrm{opt}}$ represents the
tuning function parameters which lead to minimal reconstruction error.
In this paper we focus on fixed centers of the tuning functions, and
study the optimal widths of the sensory tuning functions that minimize
$\mathrm{MMSE}(\boldsymbol{\theta})$. \\

Since it is often difficult to compute $\mathrm{MMSE}(\boldsymbol{\theta})$,
an alternative approach which has been widely used in the neural computation
literature is based on minimizing a more tractable lower bound on
the MSE, an issue which we now turn to.

\subsection{Fisher-optimal width\label{sub:Fisher-optimal-width}}

In a Bayesian context, it is well known \citep{vantrees1968dea} that
for any estimator $\hat{X},$ the MSE is lower bounded by the Bayesian
Cramér-Rao lower bound (BCRB) given in \eqref{eq:BayesianCRB}. An
interesting question is the following: considering that the MSE of
any estimator (and thus the MMSE itself) is lower bounded by the BCRB,
do the optimal widths also minimize the BCRB? In other words, can
the BCRB be used as a proxy to the MMSE in selecting optimal widths?
If this were the case, we could analyze MMSE-optimality by searching
for widths that maximize the expected value of the Fisher information
defined in \eqref{eq:FisherInformation}. This alternative analysis
would be favorable, because in most cases analytical computation of
$\mathbb{E}[\mathcal{J}(X)]$ is much simpler than that of the MMSE,
especially when the conditional likelihood is separable, in which
case the population's log likelihood reduces to a sum of individual
log likelihood functions. Moreover, it remains analytically tractable
under much broader conditions. Unfortunately, as we show below, the
answer to the above question is negative. Despite the fact that for
any encoding ensemble the BCRB lower bounds the MMSE, and even approximates
it in the asymptotic limit, the behavior of the two (as a function
of tuning function widths) is very different, as we demonstrate below.
\\

In terms of the BCRB the existence of an optimal set of widths critically
depends on the dimensionality of the environmental state. We consider
first the univariate case, and then discuss the extension to multiple
dimensions. \\

\subsubsection{The univariate case }

In the scalar case, using the tuning functions in \eqref{eq:GaussianTC},
it follows from \eqref{eq:ExpectedFisher} that it is best to employ
infinitesimally narrow tuning functions, since $\mathbb{E}[\mathcal{J}(X)]\to\infty$
most rapidly when $\alpha_{m}\to0$ for all $m$, although Fisher
information is undefined when one of the width parameters is exactly
0. A similar result was suggested previously for non-Bayesian estimation
schemes (\citet{seung1993smr,zhang1999nts,brown2006ont}), but within
the framework that was employed there the result was not valid, as
we elucidate in Section \ref{sec:Discussion-1}. Why does Fisher information
predict that {}``narrower is better''? As a tuning function narrows,
the probability that the stimulus will appear within its effective
receptive field and evoke any activity decreases, but more and more
information could be extracted about the stimulus if the cell did
emit a spike. Evidently, in the limit $\alpha_{m}\to0$ the gain in
information dominates the low probability and $\text{BCRB}\to0$;
however,\textcolor{black}{{} this is precisely the regime where the
bound vanishes and is therefore trivial. T}his result is in complete
contrast with the MSE, which is minimal for non-zero values of $\alpha$
as we show below. More disturbingly, as argued in Section \ref{sub:The-univariate-case},
the performance is in fact the worst possible when $\alpha\rightarrow0$.
We note that the inadequacy of conclusions drawn from lower bounds
was addressed using Monte Carlo simulations by \citet{bethge2002ost}.
Unfortunately, as we explain in Section \ref{sub:Bayesian-Cram=0000E9r-Rao-bound},
the asymptotic approximation used in \citet{bethge2002ost} for the
MMSE may not hold in the Bayesian setting. \\

\subsubsection{The multivariate case }

In this case, using on the $d-$dimensional Gaussian tuning functions
\eqref{eq:MultiGaussianTC} characterized by the matrices $A_{m}$
, the problem becomes more complex as is clear from \eqref{eq:Multidim-Fisher},
because the tuning functions may have different widths along different
axes or even \textcolor{black}{non-diagonal matrices. For simplicity
we focus on the case where the matrices $A_{m}$ are diagonal. By
definition, the MSE of any estimator is the trace of its error correlation
matrix, and in a}ccordance with \eqref{eq:MultiBCRB}, Fisher-optimal
tuning functions minimize the trace of $\mathbf{J}^{-1}$, which equals
the sum of its eigenvalues. If the widths in all dimensions are vanishingly
small ($A_{m}\equiv\varepsilon\mathbf{I}$ where $\varepsilon\ll1$),
$A_{m}$ becomes negligible with respect to $\Sigma_{x}$ and $\mathbb{E}[\mathcal{J}(X)]\approx\lambda_{\max}t\sum_{m}e^{-\frac{1}{2}\xi_{m}}|\Sigma_{x}|^{-\nicefrac{1}{2}}\sqrt{|A_{m}|}A_{m}^{-1}$.
In 3D and in higher dimensions the matrix $\sqrt{|A_{m}|}A_{m}^{-1}=\mathrm{diag}(\varepsilon^{d/2-1},\ldots,\varepsilon^{d/2-1})$
is close to the zero matrix, and consequently the eigenvalues of $\mathbb{E}[\mathcal{J}(X)]$
are vanishingly small. If all widths are large ($A_{m}\equiv(1/\varepsilon)\mathbf{I}$
where $\varepsilon\ll1$), $A_{m}$ dominates $\Sigma_{x}$ and $\mathbb{E}[\mathcal{J}(X)]\approx\lambda_{\max}t\sum e^{-\frac{1}{2}\xi_{m}}A_{m}^{-2}[\Sigma_{x}+(\mathbf{c}_{m}-\boldsymbol{\mu}_{x})(\mathbf{c}_{m}-\boldsymbol{\mu}_{x})^{T}]$,
namely the eigenvalues of $\mathbb{E}[\mathcal{J}(X)]$ are still
vanishingly small. This means that when the widths are too small or
too large, the BCRB dictates poor estimation performance in high dimensional
settings (specifically, the information gain no longer dominates the
low probability in the limit of vanishing widths). Therefore, there
exists an optimal set of finite positive widths that minimize the
BCRB. As an exception, in 2D the eigenvalues of $\sqrt{|A_{m}|}A_{m}^{-1}$
are finite when the widths approach 0, and $\mathbb{E}[\mathcal{J}(X)]$
cannot be said to have infinitesimally small eigenvalues. In fact,
numerical calculations for radial prior distribution reveal that in
2D {}``narrower is still better'', as in the univariate case. \\

An interesting observation concerning the BCRB based optimal widths
is that they \emph{cannot} depend on the available decoding time $t$,
since $\mathbb{E}[\mathcal{J}(X)]$ is simply proportional to $t$.
As we will show, an important consequence of the present work is to
show that the optimal tuning function widths, based on minimizing
the MMSE, depend explicitly on time. In this sense, choosing optimal
tuning functions based on the BCRB leads to a qualitatively incorrect
prediction. We return to this issue in Section \ref{sec:Discussion-1},
where further difficulties arising from the utilization of lower bounds
are presented.

\subsection{MMSE-based optimal width\label{sub:MMSE-based-optimal-width}}

Having discussed predictions about optimal widths based on BCRB minimization,
we wish to test their reliability by finding optimal widths through
direct MMSE minimization. As explained in Section \ref{sec:Methods},
in order to facilitate an analytic derivation of the MMSE, we examine
here the case of dense equally spaced tuning functions with \emph{uniform}
widths. We consider first the univariate case and then proceed to
the multivariate setting.

\subsubsection{The univariate case\label{sub:The-univariate-case}}

\begin{figure}
\begin{tabular}{ll}
A & B\tabularnewline
\includegraphics[width=85mm]{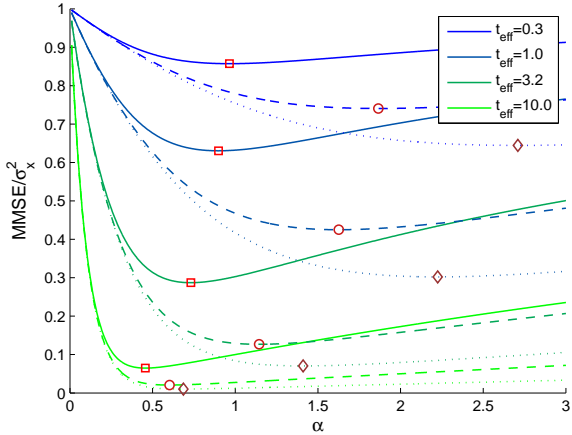} & \includegraphics[width=85mm]{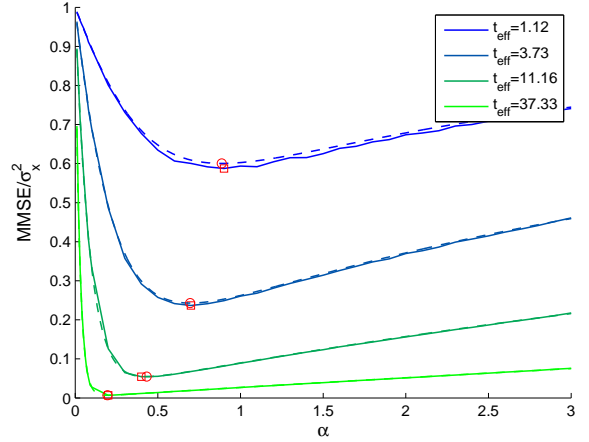}\tabularnewline
\end{tabular}

\caption{\label{Flo:MMSEvsAlpha}(A) The normalized MMSE of a dense population
of Gaussian tuning functions against tuning functions width, for different
values of effective decoding time (shortest - uppermost triplet, longest
- lowermost triplet). In each triplet 3 values of prior standard deviation
were examined - $\sigma_{x}=1$ (solid), $\sigma_{x}=2$ (dashed)
and $\sigma_{x}=3$ (dotted), where the optimal width is indicated
by squares, circles and diamonds, respectively. (B) The normalized
MMSE for different values of effective decoding time ($\sigma_{x}=1$),
obtained theoretically (dashed lines, optimal width indicated by circles)
and in simulation (solid lines, optimal width indicated by squares).}

\end{figure}

Based on \eqref{eq:MMSE_Normalized}, we plot the normalized MMSE
as a function of the width (figure \ref{Flo:MMSEvsAlpha}(A)) for
different combinations of effective decoding time and prior standard
deviation. The effective time, $t_{\text{eff}}$, defined in Section
\ref{sub:Analytical-MSE}, is proportional to the time over which
spikes accumulate. The existence of an optimal width, which is not
only positive but also varies with $\sigma_{x}$ and $t_{\text{eff}}$,
is clearly exhibited. A similar result was first demonstrated by \citet{bobrowski2009bfi},
but the dependence of optimal width on the two other parameters was
not examined. Note that the derivation in \eqref{eq:MMSE_Normalized}
is not reliable in the vicinity of the $y$-axis, because the approximation
of uniform population firing rate (see Section \eqref{sub:Analytical-MSE})
is not valid when $\alpha\to0$. Nevertheless, it can be proved that
when $\alpha\to0$, $P(\mathbf{N}_{t}=\mathbf{0})\to1$ and as a consequence
$\hat{X}^{\text{opt}}(\mathbf{N}_{t})\to\mu_{x}$ and indeed $\text{MMSE}/\sigma_{x}^{2}\to1$.\\

The existence of an optimal width is very intuitive, considering the
tradeoff between multiplicity of spikes and informativeness of observations.
When tuning functions are too wide, multiple spikes will be generated
that loosely depend on the stimulus, and thus the observations carry
little information. When tuning functions are too narrow, an observed
spike will be informative about the stimulus but the generation of
a spike is very unlikely. Thus, an intermediate optimal width emerges.
To verify that the results were not affected by the approximation
in \eqref{eq:GaussianPosterior}, we compare in figure \ref{Flo:MMSEvsAlpha}(B)
the theoretical MMSE with the MMSE obtained in simulations. \textcolor{black}{As
can be seen in the figure, the two functions coincide, and in particular,
the optimal widths are nearly identical} .\\

\begin{figure}
\begin{tabular}{ll}
A & B\tabularnewline
\includegraphics[width=85mm]{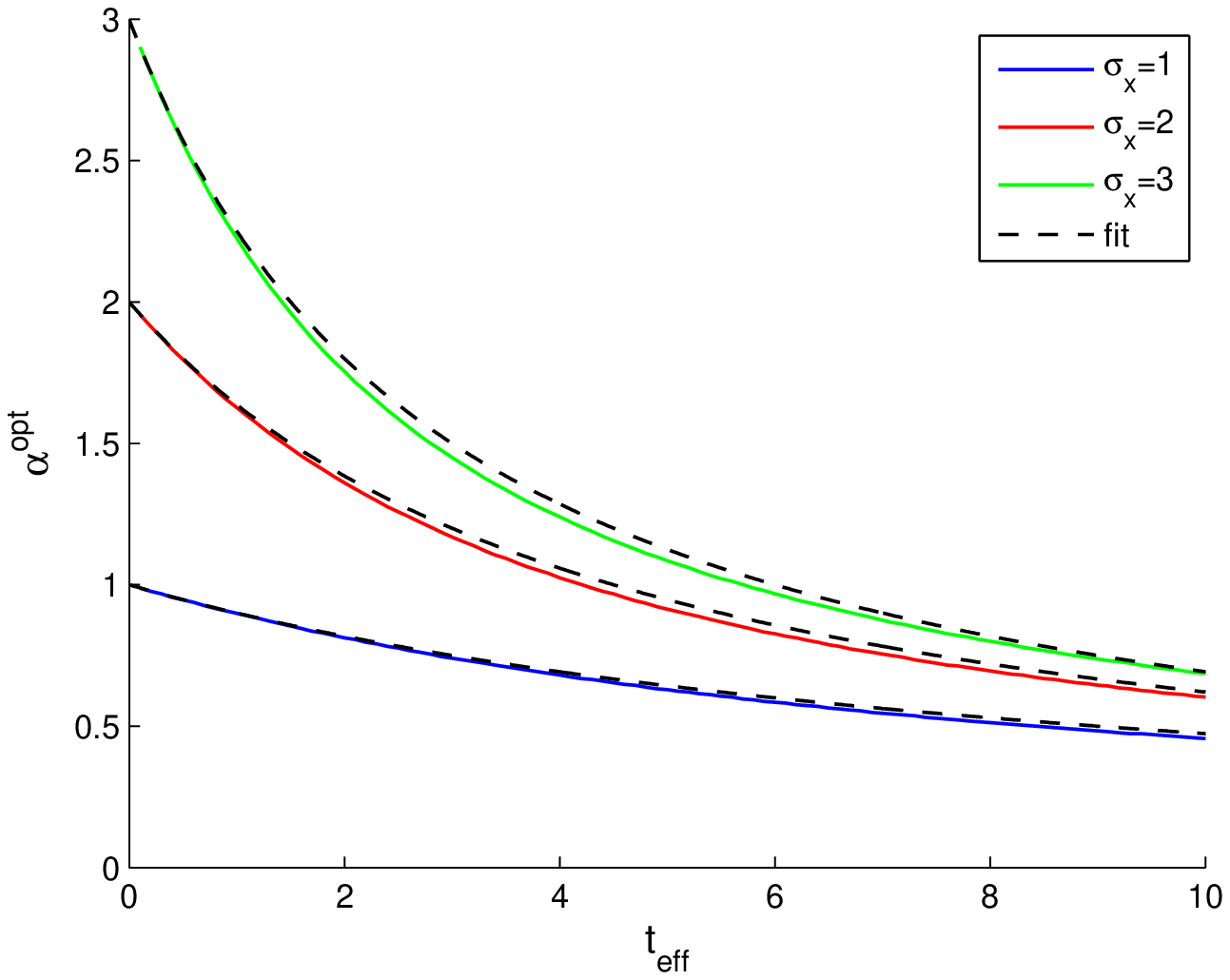} & \includegraphics[width=85mm]{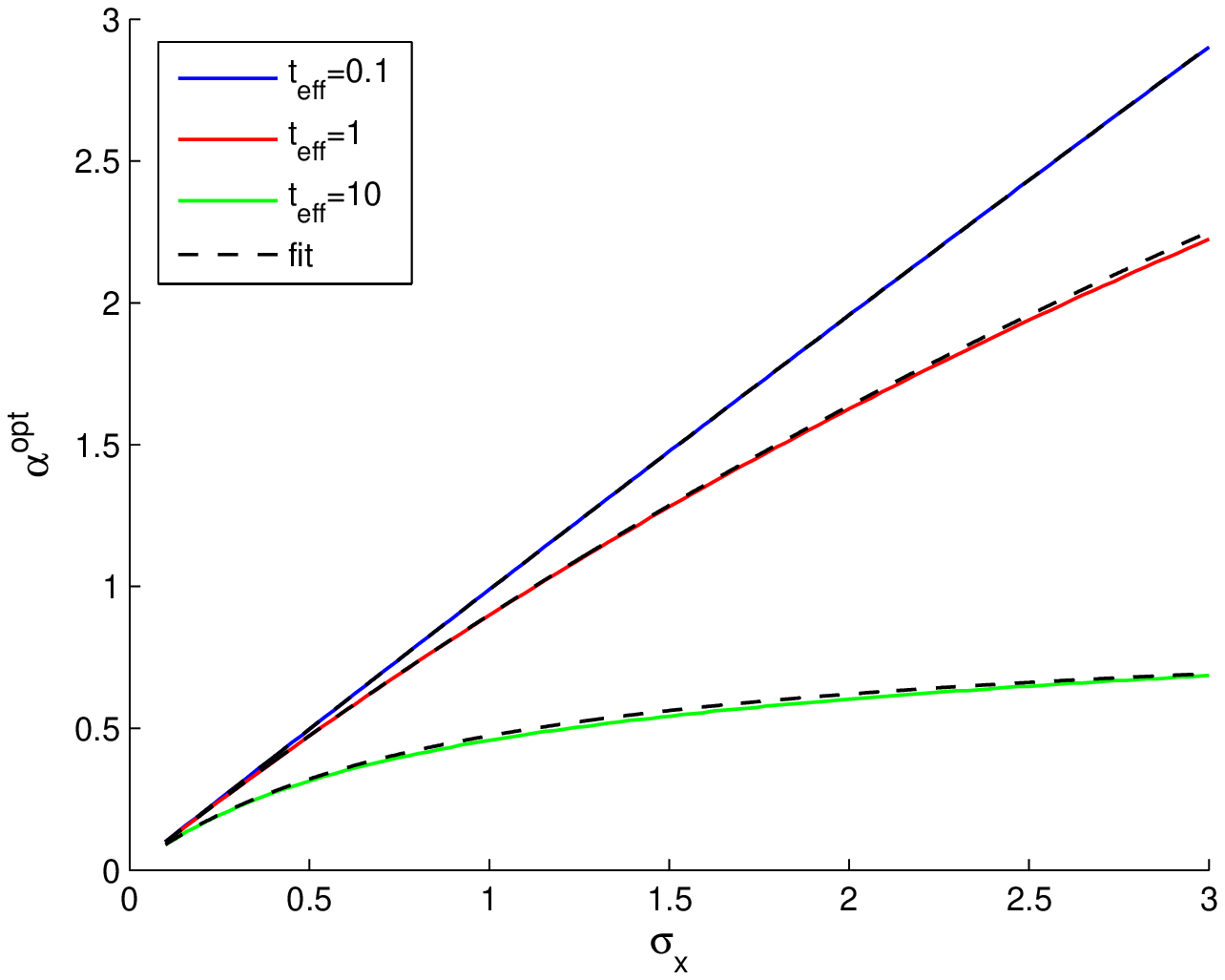}\tabularnewline
\end{tabular}

\caption{\label{Flo:AoptVsParameters}Optimal tuning functions width as a function
of (A) effective decoding time, and (B) prior standard deviation.
All curves are well fitted by the same function: $\hat{\alpha}(t_{\text{eff}},\sigma_{x})=(\frac{1}{9}t_{\text{eff}}+\sigma_{x}^{-1})^{-1}$.}

\end{figure}

We now turn to analyze the effect of the two parameters, $\sigma_{x}$
and $t_{\mathrm{eff}}$, in \eqref{eq:MMSE_Normalized}. The effective
time is a measure for the number of spikes generated by the neural
population, since it is proportional to $\lambda_{\max}t$ (where
$\lambda_{\max}$ is the maximal mean firing rate of a single neuron)
and also to $1/\Delta c$ (which indicates the population size when
the population is required to {}``cover'' a certain fraction of
the stimulus space). Longer effective times increase the likelihood
of spikes under all circumstances, and therefore the drawback of narrow
tuning functions is somewhat mitigated, leading to a reduction in
optimal width (as illustrated in figure \ref{Flo:AoptVsParameters}(A)).\\

The prior standard deviation reflects the initial uncertainty in the
environment: the larger it is - the less is known \textit{a-priori}
about the identity of the stimulus. Confidence about stimulus identity
(associated with small MMSE) may stem from either a deterministic
environment or from numerous observations for which the likelihood
is very sharp. When the environment is less certain, many more observations
are needed in order to obtain a narrow posterior distribution leading
to smaller MMSE, and thus the optimal width increases (as illustrated
in figure \ref{Flo:AoptVsParameters}(B)). Interestingly, the optimal
width is very well fitted by a simple expression of effective time
and initial uncertainty:\begin{equation}
\alpha^{\mathrm{opt}}\approx\hat{\alpha}\left(t_{\text{eff}},\sigma_{x}\right)=\left(\frac{t_{\text{eff}}}{9}+\frac{1}{\sigma_{x}}\right)^{-1}.\label{eq:OptimalWidthFit}\end{equation}

We comment in passing that the dimension of $t_{\text{eff}}$ is inverse
length. From \eqref{eq:OptimalWidthFit} we see that $\lim_{t\to0}\alpha^{\text{opt}}=\sigma_{x}$,
a result which can be obtained analytically as follows. When $t\to0$
the Poisson random variable $Y=\sum_{m}N_{t}^{m}$ converges in probability
to a Bernoulli random variable with {}``success'' probability $\alpha t_{\text{eff}}$,
in which case the expression for the normalized MMSE \eqref{eq:MMSE_Normalized}
greatly simplifies, \[
\frac{\text{MMSE}}{\sigma_{x}^{2}}=\frac{1-\alpha t_{\text{eff}}}{1}+\frac{\alpha t_{\text{eff}}}{1+\sigma_{x}^{2}/\alpha^{2}}=\frac{\alpha^{2}+\left(1-\alpha t_{\text{eff}}\right)\sigma_{x}^{2}}{\alpha^{2}+\sigma_{x}^{2}}\,.\]
Differentiating with respect to $\alpha$ and setting the result to
0 yields\[
\left[t_{\text{eff}}\sigma_{x}^{2}\left(\alpha^{2}-\sigma_{x}^{2}\right)\right]_{\alpha=\alpha^{\text{opt}}}=0\quad\Rightarrow\quad\alpha^{\text{opt}}=\sigma_{x}.\]
When the width equals the prior standard deviation, the normalized
MMSE can be calculated exactly for any effective time, \begin{eqnarray*}
\frac{\text{MMSE}}{\sigma_{x}^{2}} & = & \sum_{k=0}^{\infty}\frac{1}{1+k}\frac{\left(\sigma_{x}t_{\text{eff}}\right)^{k}e^{-\sigma_{x}t_{\text{eff}}}}{k!}=\frac{e^{-\sigma_{x}t_{\text{eff}}}}{\sigma_{x}t_{\text{eff}}}\sum_{k=0}^{\infty}\frac{\left(\sigma_{x}t_{\text{eff}}\right)^{k+1}}{\left(k+1\right)!}\\
 & = & \frac{e^{-\sigma_{x}t_{\text{eff}}}}{\sigma_{x}t_{\text{eff}}}\left(e^{\sigma_{x}t_{\text{eff}}}-1\right)=\frac{1-e^{-\sigma_{x}t_{\text{eff}}}}{\sigma_{x}t_{\text{eff}}},\end{eqnarray*}
which for very short time windows converges to the optimal normalized
MMSE.

\subsubsection{The multi-dimensional case}

We start by considering a radial prior distribution and radial tuning
functions, namely$\Sigma_{x}=\sigma_{x}^{2}\mathbf{I}$ and $A_{m}=\alpha^{2}\mathbf{I}$,
where $\mathbf{I}$ is the $d\times d$ identity matrix. In this simple
scenario the posterior covariance matrix in \eqref{eq:MultiMiuAndSigma}
becomes\[
\hat{\Sigma}=\left(\frac{1}{\sigma_{x}^{2}}+\frac{Y}{\alpha^{2}}\right)^{-1}\mathbf{I},\]
where $Y=\sum_{m}N_{t}^{m}\sim\mathrm{Pois}(\alpha^{d}t_{\text{eff}}^{d})=\mathrm{Pois}\big(\alpha t_{\text{eff}}(\sqrt{2\pi}\alpha/\Delta c)^{d-1}\big)$.
From \eqref{eq:MultiMMSE} we see that\[
\frac{\text{MMSE}}{\sigma_{x}^{2}}=d\,\mathbb{E}\left[\frac{1}{1+Y\sigma_{x}^{2}/\alpha^{2}}\right].\]
As long as $\sqrt{2\pi}\alpha^{\text{opt}}/\Delta c>1$ the parameter
of the Poisson random variable $Y$ increases with the dimensionality
$D$, in a manner that is equivalent to longer effective time (albeit
width-dependent). Consequently, recalling \eqref{eq:OptimalWidthFit},
the optimal width decreases with the dimensionality of the stimulus.
Numerical calculations show that indeed this is the case in 2D and
3D for effective times which are not extremely long. This is in contrast
with the predictions made by \citet{zhang1999nts}, where it was speculated
that performance is indifferent to the width of the tuning functions
in 2D, and improves with infinitely increasing width in higher dimensions.\\

When the prior covariance matrix and tuning functions shape matrix
$A$ are diagonal with, possibly different, elements $\{\sigma_{x,d}^{2}\}_{d=1}^{D}$
and $\{\alpha_{d}^{2}\}_{d=1}^{D}$, respectively, the MMSE is given
by\[
\text{MMSE}=\sum_{d=1}^{D}\mathbb{E}\left[\frac{1}{\sigma_{x,d}^{-2}+\alpha_{d}^{-2}Y}\right],\]
where $Y=\sum_{m}N_{t}^{m}\sim\mathrm{Pois}(t_{\text{eff}}^{D}\prod_{d=1}^{D}\alpha_{d})$.
Since the random variable $Y$ appears in every term in the sum, combined
with different prior variances, the optimal width vector will adjust
itself to the vector of prior variances, namely the optimal width
will be largest (smallest) in the dimension where the prior variance
is largest (smallest).\\

\textcolor{black}{An important feature of the theoretical results
is the ability to predict the qualitative behavior that biological
tuning functions }\textit{\textcolor{black}{should}}\textcolor{black}{{}
adopt, if they are to perform optimally. From \eqref{eq:OptimalWidthFit}
we see that if the environmental uncertainty (expressed by $\sigma_{x}$)
decreases, the tuning functions width is expected to reduce accordingly
(this holds in 2D as well). This prediction provides a possible theoretical
explanation for some results obtained in psychophysical experiments
(\citet{yeshurun1999sai}). In these experiments human subjects were
tested on spatial resolution tasks where targets appeared at random
locations on a computer screen. In each trial the subjects were instructed
to fixate on the screen center and the target was presented for a
brief moment, with or without a preceding brief spatial cue marking
the location of the target but being neutral with respect to the spatial
resolution task. The authors found that the existence of the preceding
cue improved both reaction times and success rates, concluding that
spatial resolution was enhanced following the cue by reducing the
size of neuronal receptive fields. We argue that the spatial cue reduces
the uncertainty about the stimulus by bounding the region where the
stimulus is likely to appear (i.e. $\sigma_{x}$ is reduced). In light
of our results, an optimal sensory system would then respond to the
decrease in prior standard deviation by narrowing the tuning functions
towards stimulus onset.}

\subsubsection{The effect of environmental noise}

\begin{figure}
\begin{tabular}{ll}
A & B\tabularnewline
\includegraphics[width=85mm]{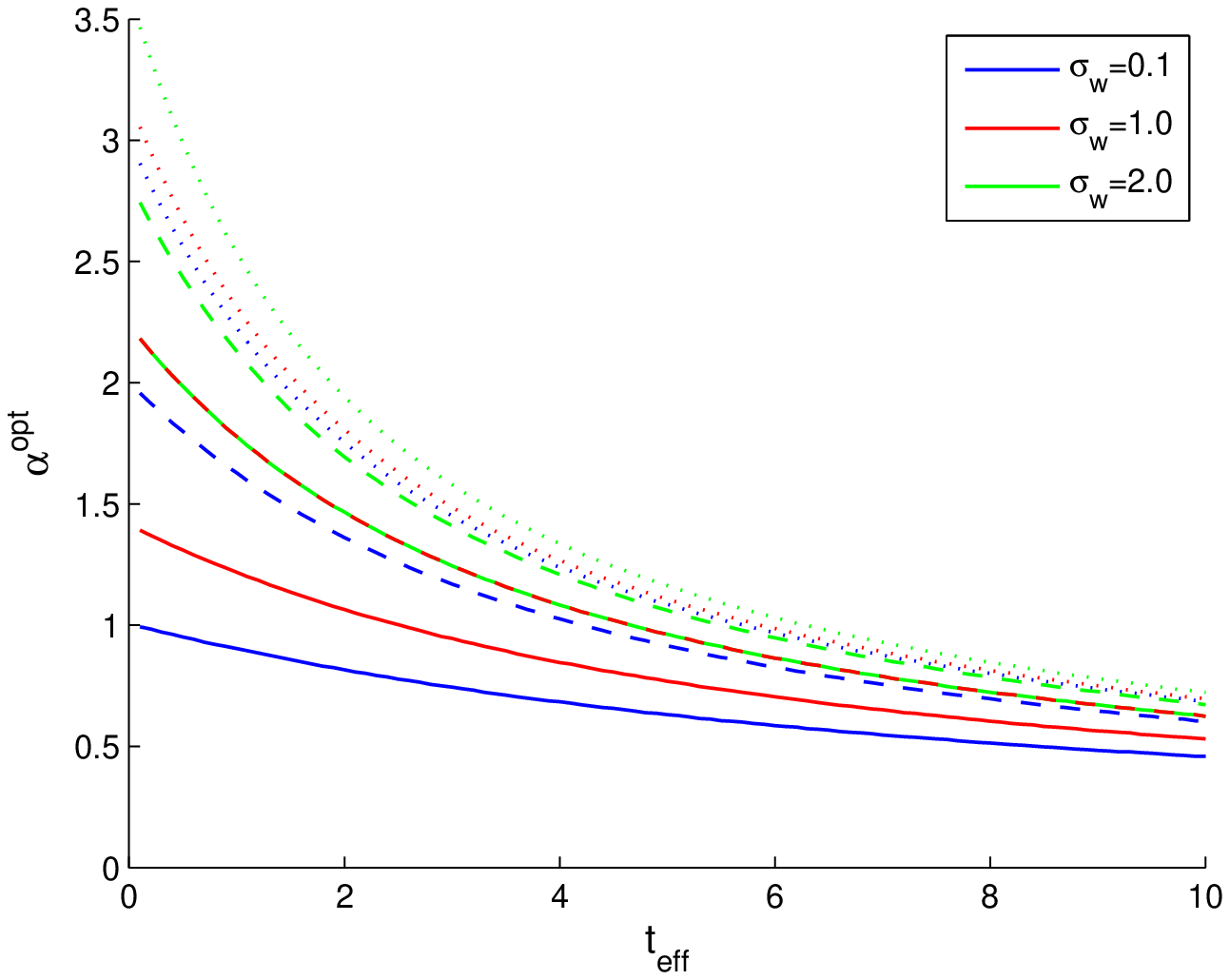} & \includegraphics[width=85mm]{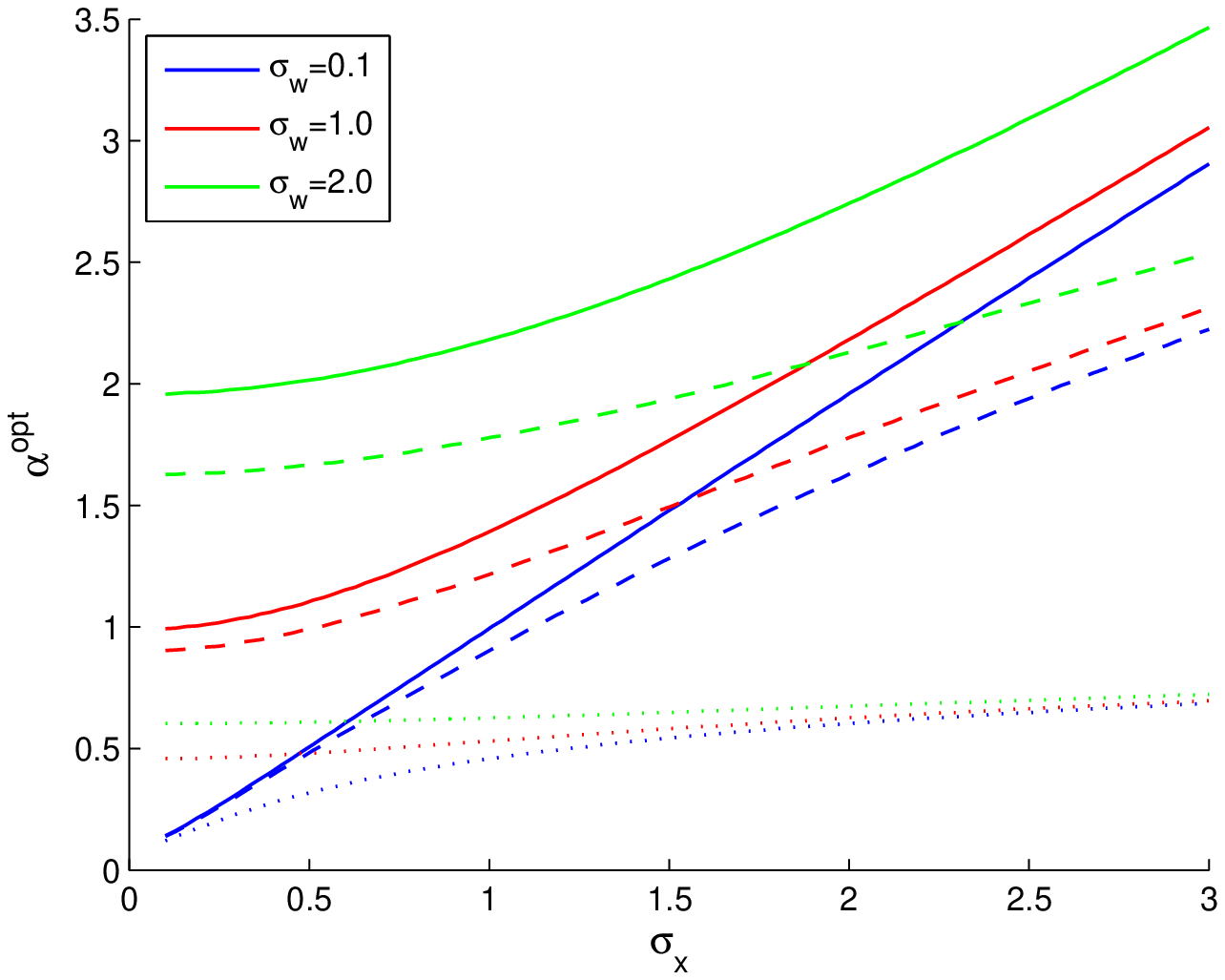}\tabularnewline
\end{tabular}

\caption{\label{Flo:AoptWithNoise}Optimal tuning functions width in the presence
of noise, for different values of parameters: (A) $\sigma_{x}=1$
(solid), $\sigma_{x}=2$ (dashed) and $\sigma_{x}=3$ (dotted); (B)
$t_{\text{eff}}=0.1$ (solid), $t_{\text{eff}}=1$ (dashed) and $t_{\text{eff}}=10$
(dotted).}

\end{figure}

In figure \ref{Flo:AoptVsParameters}(B) we saw that the optimal width
increases with $\sigma_{x}$, the prior environmental uncertainty.
Since external noise adds further uncertainty to the environment,
we expect the optimal tuning functions to broaden with increasing
noise levels, but the exact effect of noise is difficult to predict
because even in the case of additive Gaussian noise with standard
deviation $\sigma_{w}$ the expression for the MMSE \eqref{eq:NoisyMMSE}
becomes slightly more complex. The optimal width in this case is plotted
against effective time (figure \ref{Flo:AoptWithNoise}(A)) and against
prior standard deviation (figure \ref{Flo:AoptWithNoise}(B)) for
different values of noise levels. All curves are very well-fitted
by the function\[
\hat{\alpha}^{\text{opt}}\left(t_{\text{eff}},\sigma_{x}\sigma_{w}\right)=\left(\frac{t_{\text{eff}}}{9}+\frac{1}{\sqrt{\sigma_{x}^{2}+\sigma_{w}^{2}}}\right)^{-1},\]
where the average squared error of each fit is less than $1.3\times10^{-3}$.
This result implies that the effect of noise boils down to increasing
the prior uncertainty. Since the noise is additive, and is independent
of the stimulus, the term $\sqrt{\sigma_{x}^{2}+\sigma_{w}^{2}}$
is precisely the standard deviation of the noisy stimulus $\tilde{X}=X+W$,
meaning that the optimal width for estimating $X$ is the same as
for estimating $\tilde{X}$. This might not seem trivial by looking
at \eqref{eq:NoisyMMSE}, where the two standard deviations are not
interchangeable, but is nonetheless very intuitive: seeing that all
spikes depend only on $\tilde{X}$, it is the only quantity that can
be estimated directly from the spike trains, whereas $X$ is then
estimated solely from the estimator of $\tilde{X}$. Therefore, optimal
estimation of $X$ necessitates first estimating $\tilde{X}$ optimally.\\

An indication for the symmetric role played by the prior and the noise,
can be seen in figure \ref{Flo:AoptWithNoise}(A), where the dashed
red curve ($\sigma_{x}=2$, $\sigma_{w}=1$) merges with the solid
green curve ($\sigma_{x}=1$, $\sigma_{w}=2$). Note also that the
greatest effect of noise is observed when the prior standard deviation
is small, whereas for large $\sigma_{x}$ the relative contribution
of noise becomes more and more negligible (figure \ref{Flo:AoptWithNoise}(B)).

\subsubsection{The effect of multimodality}

Integration of sensory information from different modalities (e.g.
visual, auditory, somatosensory) has the advantage of increasing estimation
reliability and accuracy since each channel provides an independent
sampling of the environment (we focus here on the simple case where
tuning functions are bell-shaped in all modalities). In the absence
of noise, this improvement is merely reflected by an increased number
of observations, but the main advantage manifests itself in the noisy
case, where multimodal integration has the potential of noise reduction
since the noise variables in the two modalities are independent. Considering
two sensory modalities, denoted by $v$ and $a,$ and indexing the
respective parameters of each modality by $v$ and $a$, we provide
a closed form expression for the MMSE in \eqref{eq:BimodalMMSE-full}.
When integrating the observations, the spike trains in each modality
are weighted according to their reliability, reflected in the predictability
of the stimulus based on the estimated input (related to $\sigma_{w,v},\sigma_{w,a}$)
and in the discriminability of the tuning functions (related to $\alpha_{v},\alpha_{a}$).
For instance, when {}``visual'' noise has infinite variance or when
{}``visual'' tuning functions are flat, the spike trains in the
{}``visual'' pathway do not bear any information about the stimulus
and are thus ignored by the optimal decoder. Indeed, when substituting
$\sigma_{w,v}=\infty$ or $\alpha_{v}=\infty$ in \eqref{eq:BimodalMMSE-full},
it reduces to \eqref{eq:NoisyMMSE} with $\alpha_{a},\sigma_{w,a}$
in place of $\alpha,\sigma_{w}$. Note that in the absence of noise
the MMSE reduces to $\mathbb{E}\big[\big(\sigma_{x}^{-2}+\alpha^{-2}(Y_{v}+Y_{a})\big)^{-1}\big]$,
which is identical to the unimodal expression for the MMSE with twice
the expected number of observations ($Y_{v}+Y_{a}\sim\mathrm{Pois}(\alpha2t_{\text{eff}})$).\\

\begin{figure}
\begin{tabular}{ll}
A & B\tabularnewline
\includegraphics[width=85mm]{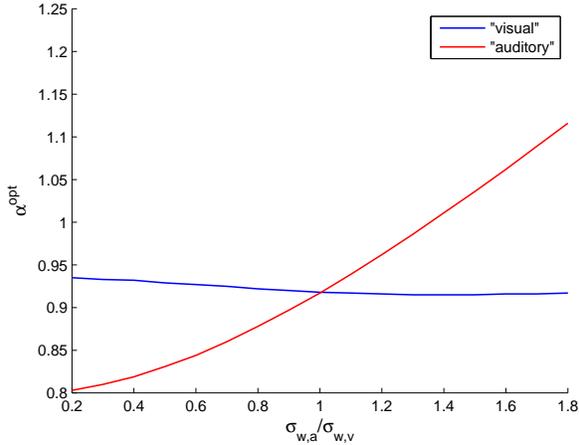} & \includegraphics[width=85mm]{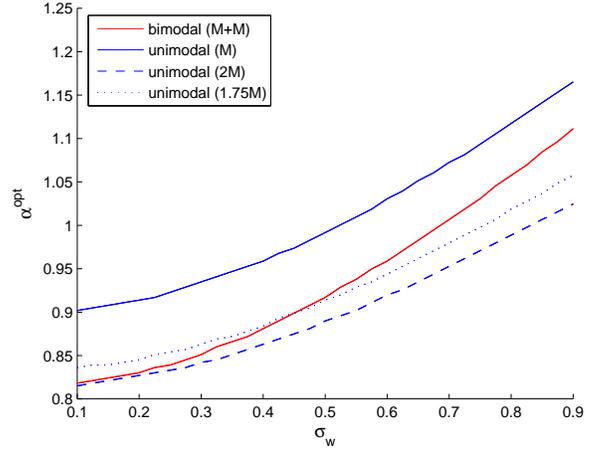}\tabularnewline
\end{tabular}

\caption{\label{Flo:MultimodalAopt}(A) Optimal tuning functions widths in
different sensory modalities as a function of the ratio between the
noise standard deviations of the underlying physical phenomena ($t_{\text{eff}}=1$,
$\sigma_{x}=1$ and $\sigma_{w,v}=0.5$). (B) Optimal tuning functions
width in symmetrical bimodal and in unimodal settings ($t_{\text{eff}}=1$
and $\sigma_{x}=1$). The numbers in parentheses denote sizes of populations.}

\end{figure}

In a bimodal setting, the optimal width in each modality adapts itself
to the standard deviation of the noise associated with that modality.
To see that, we fix $\sigma_{w,v}=0.5,$ and plot the optimal widths
against the ratio of noise standard deviations $\sigma_{w,a}/\sigma_{w,v}$
(figure \ref{Flo:MultimodalAopt}(A)). When the {}``visual'' channel
is noisier ($\sigma_{w,a}/\sigma_{w,v}<1$), the optimal {}``auditory''
tuning functions are narrower than their counterparts, and when the
{}``auditory'' channel is noisier the situation is reversed. Interestingly,
even when the {}``visual'' noise variance is fixed, the optimal
width in the {}``visual'' modality is slightly affected by the {}``auditory''
noise variance.\\

In the symmetric case we can easily compare the unimodal and bimodal
settings. Assuming that the number of tuning functions in each modality
is the same, integrating two modalities doubles the expected number
of observations, naturally resulting in a smaller optimal width (figure\ref{Flo:MultimodalAopt}(B)).
When the overall number of tuning functions is maintained constant,
splitting them into two populations of equal size in each modality
is preferred in terms of MMSE (results not shown), because the bimodal
setting also has the advantage of partial noise-cancellation. But
what can be said about the optimizing width in each setting? In figure
\ref{Flo:MultimodalAopt}(B) we observe that $2M$ optimal tuning
functions in a single modality are narrower than $M+M$ optimal tuning
functions for two different modalities. What is more intriguing is
the fact that single-modality optimal tuning functions are still narrower
than their double-modality counterparts even when they are outnumbered,
as long as the noise standard deviation is sufficiently large. \\

\textcolor{black}{The latter prediction may be related to experimental
results dealing with multisensory compensation. Neuronal tuning functions
are observed not only in the sensory areas but also in associative
areas in the cortex, where cortical maps are adaptable and not strictly
defined for a single modality. \citet{korte1993ast} found that tuning
functions in the auditory cortex of cats with early blindness are
narrower than those in normal cats. We hypothesize that this phenomenon
may subserve more accurate decoding of neural spike trains. Although
optimal auditory tuning functions are wider in the absence of visual
information (figure \ref{Flo:MultimodalAopt}(B)), our results predict
that if the blindness is compensated by an increase in the number
of auditory neurons (for instance, if some visual neurons are rewired
to respond to auditory stimuli), the auditory tuning functions will
in fact be narrower than in the case of bimodal sensory integration.
Indeed, neurobiological experiments verify that the specificity of
cortical neurons can be modified under conditions of sensory deprivation:
auditory neurons of deaf subjects react to visual stimuli (\citet{finney2001vsa})
and visual neurons of blind subjects react to sounds (\citet{weeks2000pet,rauschecker1983ace}).
We predict that even when the compensation is partial, namely when
only a portion of the visual population transforms to sound-sensitive
neurons, the optimal tuning functions in a noisy environment are still
narrower than the {}``original'' tuning functions in two functioning
modalities (figure \ref{Flo:MultimodalAopt}(B)).}

\subsection{Dynamic optimal width\label{sub:Dynamic-optimal-width}}

Two important features of the decoding process pertain to the time
available for decoding and to the prior information at hand. In principle,
we may expect that different attributes are required of the optimal
tuning functions under different conditions. We have seen in figure
\ref{Flo:AoptVsParameters} (see also \eqref{Flo:MMSEvsAlpha}) that
the optimal width increases with the initial uncertainty and decreases
with effective time. In this case the width is set in advance and
the quantity being optimized is the decoding performance at the \emph{end}
of the time window. We conclude that tuning functions should be narrower
when more knowledge about the stimulus is expected to be available
at time $t$. However, in realistic situations the decoding time may
depend on external circumstances which may not be known in advance.
It is more natural to assume that since observations accumulate over
time, the tuning functions should adapt in real time based on past
information, in such a way that performance is optimal at any time
instance, as opposed to some pre-specified time. To address this possibility,
we now allow the width to be a function of time and seek an optimal
width function. Moreover, it seems more functionally reasonable to
set the MMSE process as an optimality criterion rather than the MMSE
at an arbitrary time point. As a consequence, the optimal width function
may depend on the random accumulation of observations, namely it becomes
an optimal width process.\\

We begin by analyzing the simple case of a piecewise constant process
of the form\[
\alpha\left(t:\; i\Delta t\le t<\left(i+1\right)\Delta t\right)=\alpha_{i},\qquad\left(i=0,1,\dots\right)\]
and search for optimal (possibly random) variables $\{\alpha_{i}\}_{i=1}^{\infty}$.
At each moment, we assume that $\alpha(t)$ is large enough with respect
to $\Delta c$ so that the mean firing rate of the population, $\sum_{m}\lambda_{m}(x,t)$,
is independent of the stimulus. Therefore, the posterior distribution
is now\[
p\left(x\middle|\mathbf{N}_{t}\right)=c\exp\left\{ -\frac{1}{2}\left(\frac{\left(x-\mu_{x}\right)^{2}}{\sigma_{x}^{2}}+\sum\limits _{m=1}^{M}\sum_{i\left(m\right)=1}^{N_{t}^{m}}\frac{\left(x-c_{m}\right)^{2}}{\left[\alpha\left(t_{i\left(m\right)}\right)\right]^{2}}\right)\right\} ,\]
where $t_{i(m)}$ is the time of the $i(m)$-th spike generated by
the \textit{m}-th sensory cell. Summing over all spike-times in each
$\Delta t$-long interval, it is simple to show that at the end of
the $K$-th interval the posterior density function is Gaussian with
variance\[
\hat{\sigma}^{2}=\left(\frac{1}{\sigma_{x}^{2}}+\sum_{i=0}^{K-1}\frac{Y_{i}}{\alpha_{i}^{2}}\right)^{-1},\]
where $Y_{i}=\sum_{m}(N_{(i+1)\Delta t}^{m}-N_{i\Delta t}^{m})\sim\mathrm{Pois}(\alpha_{i}\Delta t_{\text{eff}})$
for all $i$ and $\Delta t_{\text{eff}}=\left(\sqrt{2\pi}/\Delta c\right)\lambda_{\max}\Delta t$.\\

\begin{figure}
\begin{tabular}{ll}
A & B\tabularnewline
\includegraphics[width=85mm]{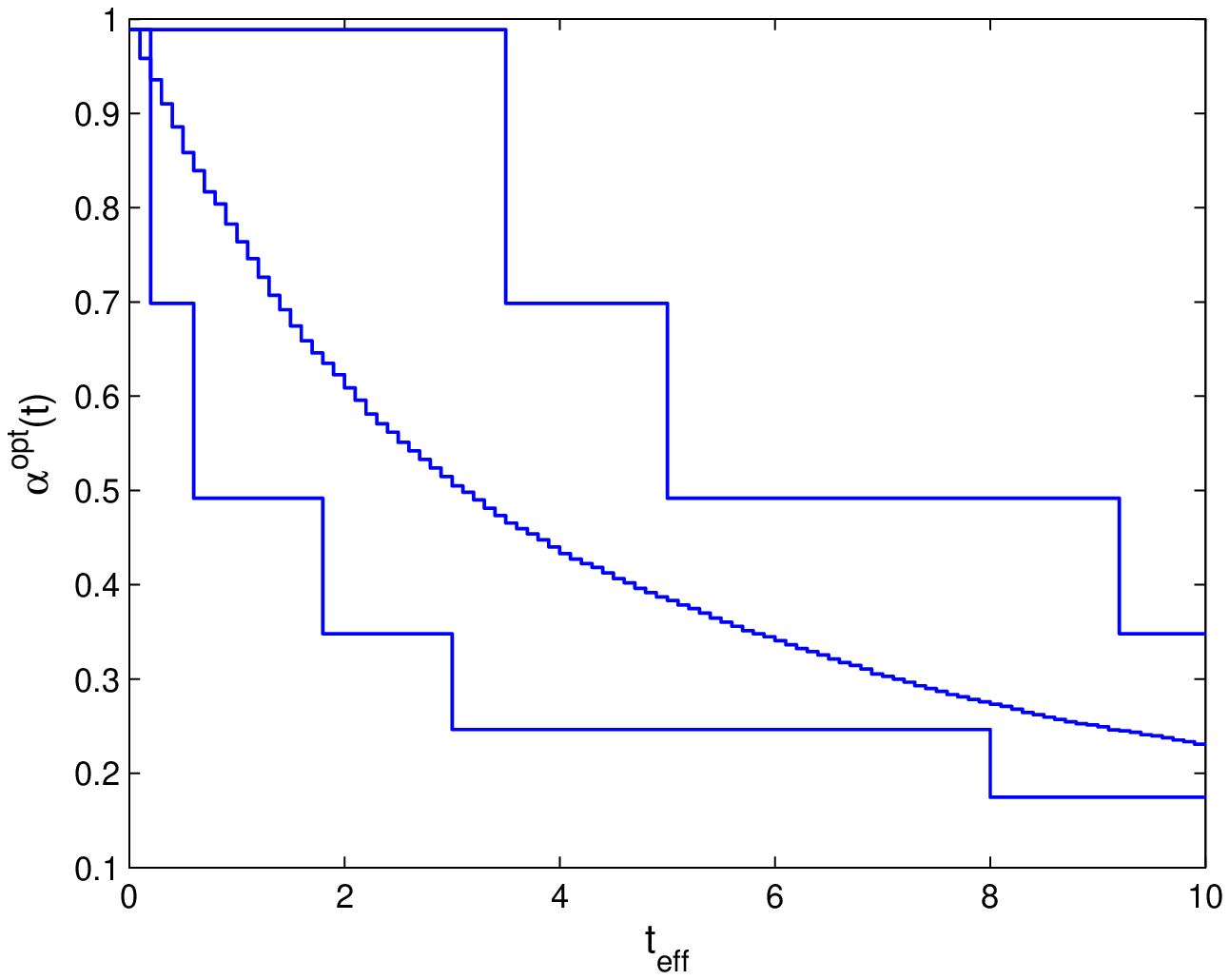} & \includegraphics[width=85mm]{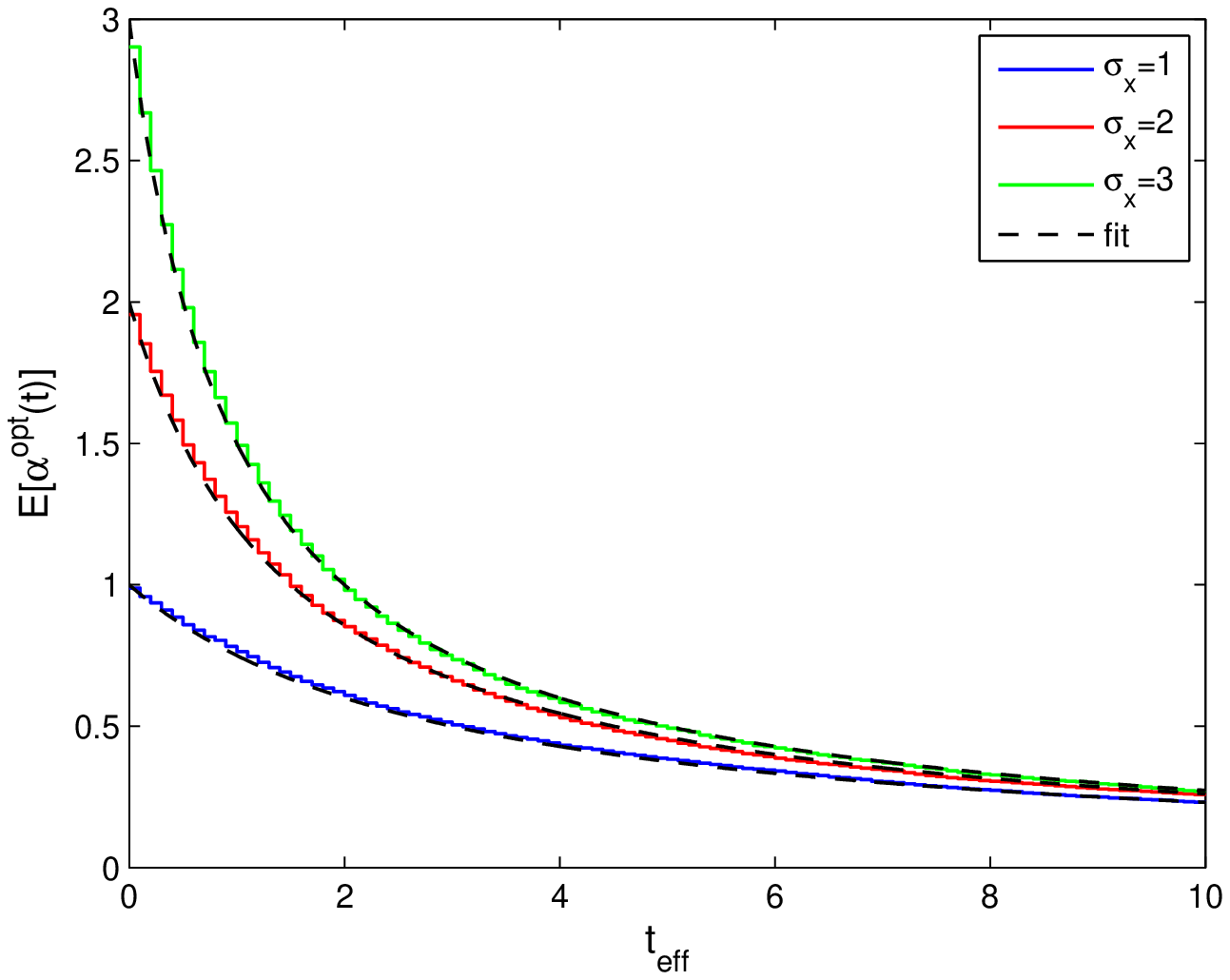}\tabularnewline
\end{tabular}

\caption{\label{Flo:OptimalProcess}(A) Sample realizations of optimal width
processes for fast (bottom) and slow (top) spiking processes. The
average over 1000 trials is plotted in the middle ($\sigma_{x}=1$
and $\Delta t_{\text{eff}}=0.1$). (B) Average optimal width processes
for different values of prior standard deviation, all coinciding with
the function $\hat{\alpha}(t,\sigma_{x})=(\frac{1}{3}t_{\text{eff}}(t)+\sigma_{x}^{-1})^{-1}$.}

\end{figure}

The value of $\alpha_{0}^{\text{opt}}$ is obtained by minimizing
$\mathbb{E}[(\sigma_{x}^{-2}+\alpha^{-2}Y_{0})^{-1}]$. When {}``choosing''
the optimal width for the second interval, the population can rely
on the spikes observed so far to modify its width so as to minimize
$\mathbb{E}[(\sigma_{1}^{-2}+\alpha^{-2}Y_{1})^{-1}]$, where $\sigma_{1}^{-2}=\sigma_{x}^{-2}+\alpha_{0}^{-2}Y_{0}$.
By recursion, at time $t=i\Delta t$ the optimal width parameter $\alpha_{i}$
is determined by taking into account all spikes in the interval $[0,(i-1)\Delta t]$
and minimizing $\mathbb{E}[(\sigma_{i}^{-2}+\alpha^{-2}Y_{i})^{-1}]$,
where\[
\sigma_{i}^{2}=\left(\frac{1}{\sigma_{x}^{2}}+\sum_{j=0}^{i-1}\frac{Y_{j}}{\alpha_{j}^{2}}\right)^{-1}\]
reflects the effective uncertainty at time $i\Delta t$ after integrating
prior knowledge and information from observations. We see that the
optimal width process is monotonically nonincreasing (since the sequence
$\{\sigma_{i}^{2}\}$ is nonincreasing), and the rate of reduction
in tuning functions width is affected by the rate of spike arrivals.
Examples for both slow and fast spiking processes can be seen in figure
\ref{Flo:OptimalProcess}(A), where the average optimal process (averaged
over $1000$ trials) is plotted as well. In general, an optimal width
process is unlikely to drastically deviate from the average due to
an internal {}``control'' mechanism: if few spikes were obtained
then the width is still relatively large, increasing the chances of
future spikes, whereas multiplicity of spikes results in small width,
limiting the probability of observing more spikes. In the limit $\Delta t\to0$,
the optimal width process starts exactly at $\sigma_{x}$ and jumps
to a value $\sqrt{2}$-times smaller than its previous value at the
occurrence of each spike (see Section S4 in the Supplementary Material
for proof).\\

When we examine the average optimal width process for different values
of prior standard deviation (figure \ref{Flo:OptimalProcess}(B)),
we see that prior to the encoding onset it is always best to keep
the tuning functions practically as wide as the probability density
function of the stimulus (i.e., at $t=0$, before any spikes are obtained,
set $\alpha^{\text{opt}}(t=0)=\sigma_{x}$ for all realizations).
The average rate of dynamic narrowing is then related to the initial
uncertainty, where the fastest narrowing occurs in the most uncertain
environment. Note that under the conditions stated in Section \ref{sub:Analytical-MSE},
the mean population activity $\sum_{m}\lambda_{m}(x)$ is proportional
to the width parameter $\alpha$. Thus, figure \ref{Flo:OptimalProcess}(B)
can be interpreted as reflecting a dynamic decrease in population
activity as a function of time. Furthermore, there is an excellent
fit between the average optimal process to the simple function\[
\hat{\alpha}\left(t;\sigma_{x}\right)=\left(\frac{t_{\text{eff}}\left(t\right)}{3}+\frac{1}{\sigma_{x}}\right)^{-1}.\]
\\

\begin{figure}
\includegraphics[width=85mm]{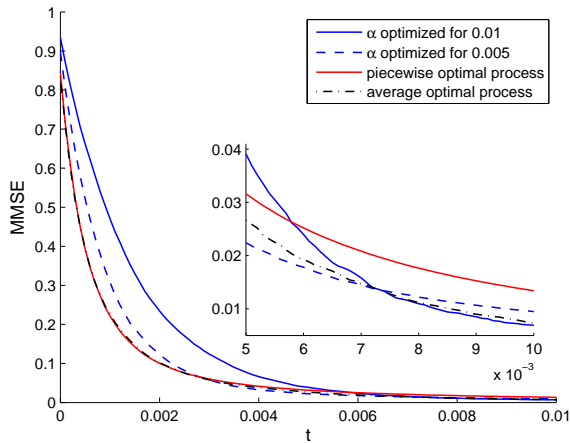}

\caption{\label{Flo:ConstantVsFunction}The MMSE associated with four different
populations, each one tuned to a different optimality criterion. The
MMSE was assessed using Monte Carlo methods. \textit{Inset:} Enlargement
for late times.}

\end{figure}

Implementation of an optimal width process seems biologically implausible,
due to its discontinuity and its perfect synchronization with the
spikes. It is more likely for biological tuning functions to adopt
the average optimal width process strategy, namely to automatically
commence a dynamic narrowing at the beginning of an encoding period,
independently of the generated spikes. But is the \textit{average}
optimal width process still more efficient than constant width? To
address this question, we simulated the dynamics of four populations:
two with constant widths tuned for optimal performance at the middle
and at the end of the decoding time window, respectively, one with
a piecewise-optimal width process and another one with a width function
corresponding to the average optimal width process. As expected, each
of the constant-width populations outperforms the other at the time
for which it is tuned (figure \ref{Flo:ConstantVsFunction}). The
dynamic-width populations significantly outperform them both for short
times, where for later times the differences become negligible. Therefore,
implementation of dynamic width offers a clear advantage over operation
with constant width, especially since there is seldom a reference
time of interest to which the tuning functions could be tuned in advance.
Interestingly, the predetermined width function is as good as the
optimal width process for short times and even outperforms it for
longer times, meaning that the performance is not being compromised
when a predetermined width function is employed in place of the optimal
width process.\textcolor{cyan}{}\\

\textcolor{black}{Our prediction that tuning functions should dynamically
narrow during the course of encoding a stimulus, coincides with some
reported experimental phenomena. \citet{wang2005sfa} recorded the
activity of auditory cortical neurons in awake marmoset monkeys in
response to a sound stimulus, which is the preferred stimulus of only
one of the neurons. When comparing the activity profiles of the two
neurons, they found that both fire intensively immediately following
stimulus presentation, but only the neuron which is tuned to the stimulus
maintained high activity throughout stimulus duration. Such behavior
is equivalent to (and can stem from) dynamic narrowing of receptive
fields. Widespread onset responses (caused by initially large receptive
fields) subserve fast detection of the occurrence of sounds, whereas
sustained firing in specific neurons facilitates extraction of the
exact features of the sound. In the context of one-dimensional tuning
functions, an illustrative example for dynamic narrowing was observed
by \citet{gutfreund2002gvi} in the external nucleus of the inferior
colliculus in anesthetized barn owls. The authors recorded from a
population of neurons tuned to a short range of inter-aural time differences
(ITD) in response to stimuli with different ITD values. While initial
responses were intensive for all stimuli, intensive ongoing responses
were restricted to stimuli within the ITD range preferred by the population,
implying a dynamic reduction in tuning function width. A similar phenomenon
was also obtained for single neurons.}

\section{\textcolor{black}{Discussion\label{sec:Discussion-1}}}

\textcolor{black}{In this paper we have studied optimal encoding of
environmental signals by a population of spiking neurons, characterized
by tuning functions which quantify the firing probability as a function
of the input. Within the framework of optimal Bayesian decoding, based
on the mean square reconstruction error criterion, we investigated
the properties of optimal encoding. Based on the well known inequality,
bounding the MMSE by the Bayesian Cramer-Rao lower bound \eqref{eq:BayesianCRB},
we tested the hypothesis that the tuning function width minimizing
the MMSE can be recovered by minimizing the (asymptotically tight)
lower bound. This hidden assumption was implied in previous studies
dealing with classical, non-Bayesian, estimators based on neural decoding
(\citet{seung1993smr,zhang1999nts,brown2006ont}). Unfortunately,
as argued in Section \ref{sec:Results}, the predictions were often
incompatible with the assumptions required in the derivation of the
bound, and thus could not be utilized to compare bound-based predictions
with results based on direct minimization of the MMSE. It should be
noted that in the non-Bayesian setting an optimal estimator does not
necessarily exist, because the performance of each estimator depends
on the stimulus, which is an unknown parameter. For instance, $\hat{X}(\mathbf{N}_{t})\equiv a$
is the optimal estimator if and only if $X=a$. The above-mentioned
studies overcame this problem by assuming that the tuning functions
are dense enough so that the population's Fisher information can be
approximated by an integral and thus becomes }\textcolor{black}{\emph{independent}}\textcolor{black}{{}
of the stimulus. They all concluded that in one dimension estimation
performance improves with narrower tuning curves. However, the underlying
assumption is valid only when the width is large enough with respect
to $\Delta c$, and obviously breaks down as $\alpha\to0$. Therefore,
values in the proximity of $\alpha=0$ are not part of the solution
space, and $\lim_{\alpha\to0}\mathcal{J}(X)$ cannot be estimated
using the suggested approximation. Moreover, realistic environments
are dynamic and it seems more reasonable to model their features as
random variables rather than unknown parameters. This means that optimal
tuning curves may be more fruitfully examined within a Bayesian framework.
Indeed, neurobiological evidence indicates that the nervous system
can adapt to changes in the statistical nature of the environment
and its random features at multiple time scales (e.g \citet{pettet1992dcr,brenner2000arm,dragoi2000aip,dean2005npc,hosoya2005dpc})}\\

Starting from the Bayesian Cramér-Rao lower bound \eqref{eq:BayesianCRB}
we have studied predictions about tuning function properties based
on this, asymptotically tight, bound. As we demonstrated, this bound-based
approach has little value in predicting the true optimal tuning functions
for finite decoding time. Even though the performance inequalities
hold in all scenarios, optimizing the bound does not guarantee the
same behavior for the bounded quantity. Moreover, an important, and
often overlooked observation is the following. Performance bounds
might be too model-dependent, and when the model is misspecified (as
is often the case) they lose their operative meaning. For instance,
when the model is incorrect, the MMSE does not converge asymptotically
to the BCRB, even though the estimator itself might converge to the
true value of the state (\citet{white1982mle}). Thus, given that
models are often inaccurate, the danger of using bounds in apparent
even in the asymptotic limit. \\

\textcolor{black}{By deriving analytical expressions for the minimal
attainable MSE under various scenarios we have obtained optimal widths
directly by minimizing the MMSE, without relying on bounds. Our analysis
uncovers the dependence of optimal width on decoding time window,
prior standard deviation and environmental noise. The relation between
optimal width and prior standard deviation may very well explain physiological
responses induced by attention, as noted in Section \ref{sub:MMSE-based-optimal-width},
where further experimental evidence supporting the predictions of
this paper was discussed. In particular, the work of \citet{korte1993ast}
related to tuning functions in the auditory cortex of cats with early
blindness was discussed, and specific predictions were made. We also
showed that when the constant-width restriction is removed, tuning
functions should narrow during the course of encoding a stimulus,
a phenomenon which has already been demonstrated for A1 neurons in
marmoset monkeys (\citet{wang2005sfa}) and in the external nucleus
of the inferior colliculus in barn owls (e.g., \citet{gutfreund2002gvi}).}\textcolor{cyan}{}\\

In summary, in this paper we examined optimal neuronal tuning functions
within a Bayesian setting based on the minimum mean square error criterion.
Careful calculations were followed by theoretical results, based on
a natural criterion of optimality which is commonly employed, but
which has seldom been analyzed in the context of neural encoding.
Our analysis yielded novel predictions about the context-dependence
of optimal widths, stating that optimal tuning curves should adapt
to the statistical properties of the environment - in accordance with
ecological theories of sensory processing. Interestingly, the results
predict at least two time scales of change. For example, when the
statistical properties of the environment change (e.g., a change in
the noise level or prior distribution) the optimal encoding should
adapt on the environmental time scale in such a way that the tuning
function widths increases with noise level. However, even in the context
of a fixed stimulus, we predict that tuning functions should change
on the fast time scale of stimulus presentation. Interestingly, recent
results, summarized by \citet{gollisch2010est}, consider contrast
adaptation in the retina taking place on a fast and slow time scale.
The fast time scale, related to ``contrast gain control'' occurs on
time scales of tens of millisecond, while the slow time-scale, lasting
many seconds, has been referred to as ``contrast adaptation''. While
our results do not directly relate to contrast adaptation, the prediction
that optimal adaptation takes place on at least two, widely separated,
time scales is promising. Moreover, the circuit mechanisms proposed
in (\citet{gollisch2010est}) may well subserve some of the computations
related to tuning function narrowing. These results, and the above-mentioned
experimentally observed phenomena, suggest that biological sensory
systems indeed adapt to changes in environmental statistics on multiple
time scales. The theory provided in this paper offers a clear functional
explanation of these phenomena, based on the notion of optimal dynamic
encoding.

\section{Methods\label{sec:Methods}}

As stated in Section \ref{sec:Results} in this paper we employ the
theoretical framework of optimal signal estimation in the context
of neural encoding and decoding, in an attempt to find the optimal
tuning functions that facilitate optimal estimation of $X$ from the
observations (the neural spike trains). We refer the reader to the
beginning of Section \ref{sec:Results} for the relevant problem setup
and definitions.

\subsection{Optimal decoding of neural spike trains}

As is well known \citep{vantrees1968dea}, the optimal estimator minimizing
the MSE is given by $\hat{X}^{\text{opt}}(\mathbf{N}_{t})=\mathbb{E}[X|\mathbf{N}_{t}]$.
This estimator can be directly computed using the posterior distribution,
obtained from Bayes' theorem\begin{equation}
p\left(x\middle|\mathbf{N}_{t}\right)=\frac{p\left(x\right)P\left(\mathbf{N}_{t}\middle|x\right)}{P\left(\mathbf{N}_{t}\right)}=cp\left(x\right)e^{-\sum_{m}\lambda_{m}\left(x\right)t}\prod_{m=1}^{M}\left(\lambda_{m}\left(x\right)t\right)^{N_{t}^{m}},\label{eq:BayesTheorem}\end{equation}
where $c$ is a normalization constant. We comment that here, and
in the sequel, we use the symbol $c$ to denote a generic normalization
constant. Since such constant will play no role in the analysis, we
do not bother to distinguish between them. For analytical tractability
we restrict ourselves to the family of Gaussian prior distributions
$X\sim p(\cdot)=\mathcal{N}(\mu_{x},\sigma_{x}^{2})$, and consider
Gaussian tuning functions\begin{equation}
\lambda_{m}(x)=\lambda_{\max}e^{-\frac{\left(x-c_{m}\right)^{2}}{2\alpha_{m}^{2}}},\qquad(m=1,\ldots,M)\label{eq:GaussianTC}\end{equation}
which in many cases set a fair approximation to biological tuning
functions (e.g. \citet{pouget2000ipp,anderson2000tcn}). From \eqref{eq:BayesTheorem}
we have \begin{equation}
p\left(x\middle|\mathbf{N}_{t}\right)=ce^{-\frac{\left(x-\mu_{x}\right)^{2}}{2\sigma_{x}^{2}}}e^{-\sum_{m}\lambda_{m}\left(x\right)t}e^{-\sum\limits _{m=1}^{M}\frac{\left(x-c_{m}\right)^{2}}{2\alpha_{m}^{2}}N_{t}^{m}}.\label{eq:PosteriorDistribution}\end{equation}
In the multi-dimensional case $\mathbf{X}\sim\mathcal{N}(\boldsymbol{\mu}_{x},\Sigma_{x})$
and the Gaussian tuning functions are of the form\begin{equation}
\lambda_{m}(x)=\lambda_{\max}e^{-\frac{1}{2}\left(\mathbf{x}-\mathbf{c}_{m}\right)^{T}A_{m}^{-1}\left(\mathbf{x}-\mathbf{c}_{m}\right)},\qquad(m=1,\ldots,M).\label{eq:MultiGaussianTC}\end{equation}

\subsection{Bayesian Cramér-Rao bound\label{sub:Bayesian-Cram=0000E9r-Rao-bound}}

The Fisher information is defined by\begin{equation}
\mathcal{J}\left(X\right)=\mathbb{E}\left[\left(\frac{\partial}{\partial X}\ln P\left(\mathbf{N}_{t}\mid X\right)\right)^{2}\right].\label{eq:FisherInformation}\end{equation}
If $X$ is deterministic, as in the classic (non-Bayesian) case, and
$B(X)\triangleq\mathbb{E}[\hat{X}-X]$ is the estimation bias, then
the error variance of \textit{any non-Bayesian} estimator $\hat{X}(\mathbf{N}_{t})$
is lower bounded by $\big(1+\frac{\partial}{\partial X}B(X)\big)^{2}\mathcal{J}^{-1}(X)$,
which is the Cramér-Rao bound (\citet{vantrees1968dea}). The extension
to the Bayesian setting, where $X$ is a random variable with probability
density function $p(\cdot)$, often referred to as the Bayesian Cramér-Rao
bound (BCRB), states that \begin{equation}
\mathbb{E}\left[\left(X-\hat{X}\right)^{2}\right]\ge\frac{1}{\mathbb{E}\left[\mathcal{J}\left(X\right)\right]+\mathcal{I}\left(p\left(x\right)\right)},\label{eq:BayesianCRB}\end{equation}
where $\mathcal{I}\big(p(x)\big)=\mathbb{E}\big[\big(\frac{\partial}{\partial X}\ln p(X)\big)^{2}\big]$
and the bounded quantity is the MSE of \textit{any} estimator (\citet{vantrees1968dea}).
Note that the expectation in \eqref{eq:BayesianCRB} is taken with
respect to both the observations $\mathbf{N_{t}}$ and the state $X$,
in contrast to the non-Bayesian case where the unknown state $X$
is deterministic. The BCRB is asymptotically tight, namely in the
limit of infinite number of observed spikes ($t\to\infty$) the MMSE
estimator satisfies equation \eqref{eq:BayesianCRB} with equality.\\

The second term in the denominator of \eqref{eq:BayesianCRB} is independent
of the tuning functions, and in the case of a univariate Gaussian
prior is given by\[
\mathcal{I}\left(p\left(x\right)\right)=\mathbb{E}\left[\frac{\left(x-\mu_{x}\right)^{2}}{\sigma_{x}^{4}}\right]=\frac{1}{\sigma_{x}^{2}}\,.\]
The expected value of the population's Fisher information (derived
in Section S1 in the Supplementary Material) is given by\begin{equation}
\mathbb{E}\left[\mathcal{J}\left(X\right)\right]=\lambda_{\text{max}}t\sum_{m=1}^{M}\frac{e^{-\frac{\left(c_{m}-\mu_{x}\right)^{2}}{2\left(\alpha_{m}^{2}+\sigma_{x}^{2}\right)}}}{\alpha_{m}\left(\alpha_{m}^{2}+\sigma_{x}^{2}\right)^{\nicefrac{5}{2}}}\left[\sigma_{x}^{2}\left(\sigma_{x}^{2}+\alpha_{m}^{2}\right)+\alpha_{m}^{2}\left(c_{m}-\mu_{x}\right)^{2}\right].\label{eq:ExpectedFisher}\end{equation}
\\

In the multi-dimensional case the right hand side of \eqref{eq:BayesianCRB}
is replaced by the inverse of the matrix $\mathbf{J}=\mathbb{E}[\mathcal{J}(\mathbf{X})]+\mathcal{I}\big(p(\mathbf{x})\big)$,
where\begin{eqnarray*}
\mathcal{J}(\mathbf{X}) & = & \mathbb{E}\left[\left(\nabla\ln P\left(\mathbf{N}_{t}\mid\mathbf{X}\right)\right)\left(\nabla\ln P\left(\mathbf{N}_{t}\mid\mathbf{X}\right)\right)^{T}\right],\\
\mathcal{I}\left(p\left(\mathbf{x}\right)\right) & = & \mathbb{E}\left[\left(\nabla\ln P\left(\mathbf{X}\right)\right)\left(\nabla\ln P\left(\mathbf{X}\right)\right)^{T}\right],\end{eqnarray*}
and the left hand side is replaced by the error correlation matrix
$\mathbf{R}=\mathbb{E}\big[(\mathbf{X}-\hat{\mathbf{X}})(\mathbf{X}-\hat{\mathbf{X}})^{T}\big]$.
The interpretation of the resulting inequality is twofold. First,
the matrix $\mathbf{R}-\mathbf{J}^{-1}$ is nonnegative definite and
second, \begin{equation}
\mathbb{E}\left[\left(\mathbf{X}_{k}-\hat{\mathbf{X}}_{k}\right)^{2}\right]\ge\left(\mathbf{J}^{-1}\right)_{kk},\qquad(k=1,\ldots,d),\label{eq:MultiBCRB}\end{equation}
where $d$ is the dimension of $\mathbf{X}$ (\citet{vantrees1968dea}).\\

It is straightforward to show that in this case \[
\mathcal{I}\left(p\left(\mathbf{x}\right)\right)=\mathbb{E}\left[\Sigma_{x}^{-1}\left(\mathbf{X}-\boldsymbol{\mu}_{x}\right)\left(\mathbf{X}-\boldsymbol{\mu}_{x}\right)^{T}\Sigma_{x}^{-1}\right]=\Sigma_{x}^{-1}.\]
The expression for $\mathbb{E}\left[\mathcal{J}\left(\mathbf{X}\right)\right]$
is similar to \eqref{eq:ExpectedFisher} and is derived in Section
S1 in the Supplementary Material, yielding \begin{eqnarray}
\mathbb{E}\left[\mathcal{J}\left(\mathbf{X}\right)\right] & = & \lambda_{\text{max}}t\sum_{m=1}^{M}e^{-\frac{1}{2}\xi_{m}}\frac{\sqrt{\left|A_{m}\right|}A_{m}^{-1}}{\sqrt{\left|A_{m}+\Sigma_{x}\right|}}\left(A_{m}+\Sigma_{x}\right)^{-2}\nonumber \\
 & \times & \left[\Sigma_{x}\left(A_{m}+\Sigma_{x}\right)+A_{m}\left(\mathbf{c}_{m}-\boldsymbol{\mu}_{x}\right)\left(\mathbf{c}_{m}-\boldsymbol{\mu}_{x}\right)^{T}\right]\,,\label{eq:Multidim-Fisher}\end{eqnarray}
where $\xi_{m}=(\mathbf{c}_{m}-\boldsymbol{\mu}_{x})^{T}(A_{m}+\Sigma_{x})^{-1}(\mathbf{c}_{m}-\boldsymbol{\mu}_{x})$.
\\

As noted in Section \ref{sub:Fisher-optimal-width}, \citet{bethge2002ost}
were the first to address the shortcomings of lower bounds on the
MSE in a Bayesian setting. These authors stressed the fact that performance
bounds are usually tight only in the limit of infinitely long time
windows, and cannot be expected to reflect the behavior of the bounded
quantities for finite decoding times. They observed that in the classical
approach, for any $X$, the conditional MMSE for unbiased estimators
asymptotically equals $1/\mathcal{J}(X)$. By taking expectations
on both sides they obtained $\mathbb{E}[1/\mathcal{J}(X)]$ as the
asymptotic Bayesian MMSE. Unfortunately, as noted following \eqref{eq:BayesianCRB},
the asymptotically tight lower bound in the Bayesian setting is $\left[\mathbb{E}\left[\mathcal{J}\left(X\right)\right]+\mathcal{I}\left(p\left(x\right)\right)\right]^{-1}$
which can easily be seen, using Jensen's inequality, to be smaller
than $\mathbb{E}[1/\mathcal{J}(X)]$. The reason for this subtle discrepancy
is that the optimal Bayesian estimator can make use of the prior and
may be conditionally biased, while any non-Bayesian estimator does
not have access to the prior distribution. Thus, the Bayesian error
may be smaller than $\mathbb{E}[1/\mathcal{J}(X)]$ \\

\subsection{Analytical derivation of the MMSE\label{sub:Analytical-MSE}}

In this Section we proceed to establish closed form expressions for
the MMSE. In order to maintain analytical tractability we analyze
the simple case of equally spaced tuning functions ($c_{m+1}-c_{m}\equiv\Delta c$)
with uniform width ($\alpha_{m}\equiv\alpha$). When the width is
of the order of $\Delta c$ or higher, and $M$ is large enough with
respect to $\sigma_{x}$, the mean firing rate of the entire population
is practically uniform for any {}``reasonable'' $X$. As shown in
Section S2 in the Supplementary Material, the sum $\sum_{m}\lambda_{m}(x)$
is well approximated in this case by $\lambda(\alpha)=\sqrt{2\pi}\frac{\alpha}{\Delta c}\lambda_{\max}$.
Under these conditions the posterior \eqref{eq:PosteriorDistribution}
takes the form \begin{equation}
p\left(x\middle|\mathbf{N}_{t}\right)=c\exp\left\{ -\frac{1}{2}\left(\frac{\left(x-\mu_{x}\right)^{2}}{\sigma_{x}^{2}}+\sum\limits _{m=1}^{M}\frac{\left(x-c_{m}\right)^{2}}{\alpha^{2}}N_{t}^{m}\right)\right\} ,\label{eq:GaussianPosterior}\end{equation}
implying that $X|\mathbf{N}_{t}\sim\mathcal{N}(\hat{\mu},\hat{\sigma}^{2})$,
where\[
\hat{\sigma}^{2}=\left(\frac{1}{\sigma_{x}^{2}}+\frac{\sum_{m}N_{t}^{m}}{\alpha^{2}}\right)^{-1}\quad;\quad\hat{\mu}=\hat{\sigma}^{2}\left(\frac{\mu_{x}}{\sigma_{x}^{2}}+\frac{\sum_{m}c_{m}N_{t}^{m}}{\alpha^{2}}\right).\]
Recalling that the spike trains are independent Poisson processes,
namely $\sum_{m}N_{t}^{m}|X\sim\mathrm{Pois}\big(\sum_{m}\lambda_{m}(X)t\big)$,
and in light of the above approximation, $Y\triangleq\sum_{m}N_{t}^{m}$
is independent of $X$ and $Y\sim\mathrm{Pois}(\alpha t_{\text{eff}})$,
where $t_{\text{eff}}\triangleq\frac{\sqrt{2\pi}}{\Delta c}\lambda_{\max}t$.
Since $\hat{X}^{\text{opt}}(\mathbf{N}_{t})=\mathbb{E}[X|\mathbf{N}_{t}]\equiv\hat{\mu}$,
the MMSE is given by \[
\text{MMSE}=\mathbb{E}\left[\left(X-\hat{X}^{\text{opt}}\right)^{2}\right]=\mathbb{E}\left[\mathbb{E}\left[\left(X-\hat{\mu}\right)^{2}\middle|\mathbf{N}_{t}\right]\right]\equiv\mathbb{E}\left[\hat{\sigma}^{2}\right],\]
and thus we get an explicit expression for the normalized MMSE, \begin{equation}
\frac{\text{MMSE}}{\sigma_{x}^{2}}=\mathbb{E}\left[\frac{1}{1+Y\sigma_{x}^{2}/\alpha^{2}}\right]=\sum_{k=0}^{\infty}\frac{1}{1+k\sigma_{x}^{2}/\alpha^{2}}\frac{\left(\alpha t_{\text{eff}}\right)e^{-\alpha t_{\text{eff}}}}{k!}.\label{eq:MMSE_Normalized}\end{equation}
In the multivariate case we assume that the centers of the tuning
functions form a dense multi-dimensional lattice with equal spacing
$\Delta c$ along all axes, and \eqref{eq:GaussianPosterior} becomes
\begin{equation}
p\left(\mathbf{x}\middle|\mathbf{N}_{t}\right)=c\exp\left\{ -\frac{1}{2}\left(\left(\mathbf{x}-\boldsymbol{\mu}_{x}\right)^{T}\Sigma_{x}^{-1}\left(\mathbf{x}-\boldsymbol{\mu}_{x}\right)+\sum\limits _{m=1}^{M}N_{t}^{m}\left(\mathbf{x}-\mathbf{c}_{m}\right)^{T}A^{-1}\left(\mathbf{x}-\mathbf{c}_{m}\right)\right)\right\} ,\label{eq:p(x|Nt)_multi}\end{equation}
namely $\mathbf{X}|\mathbf{N}_{t}\sim\mathcal{N}(\hat{\boldsymbol{\mu}},\hat{\Sigma})$,
where\begin{equation}
\hat{\Sigma}=\left(\Sigma_{x}^{-1}+A^{-1}\sum\nolimits _{m}N_{t}^{m}\right)^{-1};\qquad\hat{\boldsymbol{\mu}}=\hat{\Sigma}\left(\Sigma_{x}^{-1}\boldsymbol{\mu}_{x}+A^{-1}\sum\nolimits _{m}\mathbf{c}_{m}N_{t}^{m}\right).\label{eq:MultiMiuAndSigma}\end{equation}
 In this case $Y\triangleq\sum_{m}N_{t}^{m}\sim\mathrm{Pois}(\sqrt{|A|}t_{\text{eff}}^{d})$,
$t_{\text{eff}}^{d}\triangleq\big(\sqrt{2\pi}/\Delta c\big)^{d}\lambda_{\max}t$
(see Section S2 in the Supplementary Material), and\begin{eqnarray}
\text{MMSE} & = & \mathbb{E}\left[\left\Vert \mathbf{X}-\hat{\mathbf{X}}^{\text{opt}}\right\Vert ^{2}\right]=\mathbb{E}\left[\mathbb{E}\left[\left(\mathbf{X}-\hat{\boldsymbol{\mu}}\right)^{T}\left(\mathbf{X}-\hat{\boldsymbol{\mu}}\right)\middle|\mathbf{N}_{t}\right]\right]\nonumber \\
 & = & \mathbb{E}\left[\mathbb{E}\left[\sum_{k=1}^{d}\left(\mathbf{X}_{k}-\hat{\boldsymbol{\mu}}_{k}\right)^{2}\middle|\mathbf{N}_{t}\right]\right]=\sum_{k=1}^{d}\mathbb{E}\left[\hat{\Sigma}_{kk}\right].\label{eq:MultiMMSE}\end{eqnarray}

\subsection{Incorporation of noise \label{sub:Meth-noise}}

Suppose that due to environmental noise the value of $X$ is not directly
available to the sensory system, which must respond to a noisy version
of the environmental state - $\tilde{X}$. For simplicity we assume
additive Gaussian noise: $\tilde{X}=X+W,\; W\sim\mathcal{N}(0,\sigma_{w}^{2})$.
For a dense layer of sensory neurons the joint posterior distribution
of the environmental state and noise is\begin{eqnarray*}
p\left(x,w\middle|\mathbf{N}_{t}\right) & = & \frac{p\left(x,w\right)P\left(\mathbf{N}_{t}\middle|x,w\right)}{P\left(\mathbf{N}_{t}\right)}=\frac{p\left(x\right)p\left(w\right)P\left(\mathbf{N}_{t}\middle|x+w\right)}{P\left(\mathbf{N}_{t}\right)}\\
 & = & ce^{-\frac{\left(x-\mu_{x}\right)^{2}}{2\sigma_{x}^{2}}}e^{-\frac{w^{2}}{2\sigma_{w}^{2}}}e^{-\sum\limits _{m=1}^{M}\frac{\left(x+w-c_{m}\right)^{2}}{2\alpha^{2}}N_{t}^{m}}\\
 & = & ce^{-\frac{1}{2\hat{\sigma}_{w}^{2}}\left(w-\hat{\sigma}_{w}^{2}\sum\limits _{m=1}^{M}\frac{\left(c_{m}-x\right)}{\alpha^{2}}N_{t}^{m}\right)^{2}}e^{-\frac{1}{2}\left\{ \frac{\left(x-\mu_{x}\right)^{2}}{\sigma_{x}^{2}}+\sum\limits _{m=1}^{M}\frac{\left(x-c_{m}\right)^{2}}{\alpha^{2}}N_{t}^{m}-\hat{\sigma}_{w}^{2}\left(\sum\limits _{m=1}^{M}\frac{\left(c_{m}-x\right)}{\alpha^{2}}N_{t}^{m}\right)^{2}\right\} },\end{eqnarray*}
which is Gaussian with respect to $W$, where $\hat{\sigma}_{w}^{2}=(\sigma_{w}^{-2}+\alpha^{-2}\sum_{m}N_{t}^{m})^{-1}$.
Integration over $w$ gives the marginal distribution\[
p\left(x\middle|\mathbf{N}_{t}\right)=ce^{-\frac{1}{2}\left\{ \frac{\left(x-\mu_{x}\right)^{2}}{\sigma_{x}^{2}}+\sum\limits _{m=1}^{M}\frac{\left(x-c_{m}\right)^{2}}{\alpha^{2}}N_{t}^{m}-\hat{\sigma}_{w}^{2}\left(\sum\limits _{m=1}^{M}\frac{\left(c_{m}-x\right)}{\alpha^{2}}N_{t}^{m}\right)^{2}\right\} },\]
which is Gaussian with variance\begin{equation}
\hat{\sigma}^{2}=\left(\frac{1}{\sigma_{x}^{2}}+\frac{Y}{\alpha^{2}}-\frac{\hat{\sigma}_{w}^{2}Y^{2}}{\alpha^{4}}\right)^{-1}=\left(\frac{1}{\sigma_{x}^{2}}+\frac{Y}{\alpha^{2}}-\frac{\sigma_{w}^{2}Y^{2}}{\alpha^{2}\left(\alpha^{2}+\sigma_{w}^{2}Y\right)}\right)^{-1}.\label{eq:NoisySigma}\end{equation}
where $Y=\sum_{m}N_{t}^{m}\sim\mathrm{Pois}(\alpha t_{\text{eff}})$
as before. The MMSE is computed directly from the posterior variance:\begin{equation}
\text{MMSE}=\mathbb{E}\left[\hat{\sigma}^{2}\right]=\mathbb{E}\left[\left(\frac{1}{\sigma_{x}^{2}}+\frac{Y}{\alpha^{2}+\sigma_{w}^{2}Y}\right)^{-1}\right].\label{eq:NoisyMMSE}\end{equation}
Note that when $\sigma_{w}\to\infty$ the sensory information becomes
irrelevant for estimating $X$, and indeed in such case $\text{MMSE}\to\sigma_{x}^{2}$.

\subsection{Multisensory integration \label{sub:Meth-Multisensory-integration} }

Consider the case where the environmental state is estimated based
on observations from two different sensory modalities (e.g., visual
and auditory). The spike trains in each channel are generated in a
similar manner to the unimodal case, but are then integrated to yield
enhanced neural decoding. All quantities that are related to the first
and second modalities will be indexed by $v$ and $a$, respectively,
although these modalities are not necessarily the visual and the auditory.

Each sensory modality detects a different noisy version of the environmental
state: $\tilde{X}_{v}=X+W_{v}$ in the {}``visual'' pathway and
$\tilde{X}_{a}=X+W_{a}$ in the {}``auditory'' pathway. The noise
variables are Gaussian with zero-mean and variances $\sigma_{w,v}^{2}$
and $\sigma_{w,a}^{2}$, respectively, and independent of each other
as they emerge from different physical processes. The noisy versions
of the environmental state are encoded by $M_{v}+M_{a}$ sensory neurons
with conditional Poisson statistics:\[
\begin{cases}
N_{t}^{m,v}|\tilde{X}_{v}\sim\mathrm{Pois}\left(\lambda_{m}^{v}(\tilde{X}_{v})t\right), & \;(m=1,\ldots,M_{v})\\
N_{t}^{m,a}|\tilde{X}_{a}\sim\mathrm{Pois}\left(\lambda_{m}^{a}(\tilde{X}_{a})t\right), & \;(m=1,\ldots,M_{a}),\end{cases}\]
where the tuning functions in each modality are Gaussian with uniform
width, \[
\begin{cases}
\lambda_{m}^{v}\left(x\right)=\lambda_{\max}^{v}e^{-\frac{\left(x-c_{m}^{v}\right)^{2}}{2\alpha_{v}^{2}}}, & \;(m=1,\ldots,M_{v})\\
\lambda_{m}^{a}\left(x\right)=\lambda_{\max}^{a}e^{-\frac{\left(x-c_{m}^{a}\right)^{2}}{2\alpha_{a}^{2}}}, & \;(m=1,\ldots,M_{a}).\end{cases}\]
We assume that the tuning functions are dense enough ($\alpha_{v}\not\ll\Delta c_{v},\;\alpha_{a}\not\ll\Delta c_{a}$)
and equally spaced ($c_{m+1}^{v}-c_{m}^{v}\equiv\Delta c_{v},\; c_{m+1}^{a}-c_{m}^{a}\equiv\Delta c_{a}$
).

The full expression for \textcolor{black}{\large the} MMSE in this
case is

\begin{equation}
\text{MMSE}=\mathbb{E}\left[\left(\frac{1}{\sigma_{x}^{2}}+\frac{Y_{v}}{\alpha_{v}^{2}+\sigma_{w,v}^{2}Y_{v}}+\frac{Y_{a}}{\alpha_{a}^{2}+\sigma_{w,a}^{2}Y_{a}}\right)^{-1}\right],\label{eq:BimodalMMSE-full}\end{equation}
where $Y_{v}\triangleq\sum_{m}N_{t}^{m,v}\sim\text{Pois}(\alpha_{v}t_{\text{eff}}^{v})$
and $Y_{a}\triangleq\sum_{m}N_{t}^{m,a}\sim\text{Pois}(\alpha_{a}t_{\text{eff}}^{a})$
(the derivation appears in Section S3 in the Supplementary Material).\\
\bigskip{}

\subsubsection*{Disclosure/Conflict-of-Interest Statement}

The authors declare that the research was conducted in the absence
of any commercial or financial relationships that could be construed
as potential conflict of interest.\\

\subsubsection*{Acknowledgments}

This work was partially supported by grant No. 665/08 from the Israel
Science Foundation. We thank Eli Nelken and Yoram Gutfreund for helpful
discussions. \\

\pagebreak

\part*{Supplementary Material}

\section*{S1. Expectation of population Fisher information}

In this section we calculate the expected value of the Fisher information
for a population of sensory neurons following Poisson spiking statistics,
having Gaussian tuning functions of the form\[
\lambda_{m}\left(x\right)=\lambda_{\max}e^{-\frac{\left(x-c_{m}\right)^{2}}{2\alpha_{m}^{2}}},\qquad\left(m=1,\ldots,M\right)\]
and encoding a stimulus with Gaussian prior distribution ($X\sim p(\cdot)=\mathcal{N}(\mu_{x},\sigma_{x}^{2})$).
We start by calculating the population's local Fisher information,
taking into account the conditional independence of the Poisson processes:\begin{eqnarray*}
\mathcal{J}\left(X\right) & = & \mathbb{E}\left[\left(\frac{\partial}{\partial X}\ln\left(\prod_{m=1}^{M}P\left(N_{t}^{m}\middle|X\right)\right)\right)^{2}\right]=\mathbb{E}\left[\left(\sum_{m=1}^{M}\frac{\partial}{\partial X}\ln P\left(N_{t}^{m}\middle|X\right)\right)^{2}\right]\\
 & = & \mathbb{E}\left[\left(\sum_{m=1}^{M}\left(\frac{N_{t}^{m}}{\lambda_{m}\left(X\right)}-t\right)\frac{\partial\lambda_{m}\left(X\right)}{\partial X}\right)^{2}\right]=\mathbb{E}\left[\left(\sum_{m=1}^{M}\left(N_{t}^{m}-\lambda_{m}\left(X\right)t\right)\left(\frac{c_{m}-X}{\alpha_{m}^{2}}\right)\right)^{2}\right]\\
 & = & \sum_{m=1}^{M}\sum_{n=1}^{M}\mathbb{E}\left[\left(N_{t}^{m}-\lambda_{m}\left(X\right)t\right)\left(N_{t}^{n}-\lambda_{n}\left(X\right)t\right)\middle|X\right]\left(\frac{c_{m}-X}{\alpha_{m}^{2}}\right)\left(\frac{c_{n}-X}{\alpha_{n}^{2}}\right).\end{eqnarray*}
Due to the conditional independence the expectation is separable for
every $m\not=n$, and equals 0 because $\mathbb{E}[N_{t}^{m}|X]=\lambda_{m}(X)t$.
Thus,\[
\mathcal{J}\left(X\right)=\sum_{m=1}^{M}\overset{\text{var}\left(N_{t}^{m}\middle|X\right)=\lambda_{m}\left(X\right)t}{\overbrace{\mathbb{E}\left[\left(N_{t}^{m}-\lambda_{m}\left(X\right)t\right)^{2}\middle|X\right]}}\frac{\left(X-c_{m}\right)^{2}}{\alpha_{m}^{4}}=\lambda_{\text{max}}t\sum_{m=1}^{M}\frac{\left(X-c_{m}\right)^{2}}{\alpha_{m}^{4}}e^{-\frac{\left(X-c_{m}\right)^{2}}{2\alpha_{m}^{2}}}.\]
Taking the expectation over the prior $p(x)$ yields:

\begin{eqnarray*}
\mathbb{E}\left[\mathcal{J}\left(X\right)\right] & = & \int\limits _{-\infty}^{\infty}\lambda_{\text{max}}t\sum_{m=1}^{M}\left(\frac{\left(x-c_{m}\right)^{2}}{\alpha_{m}^{4}}e^{-\frac{\left(x-c_{m}\right)^{2}}{2\alpha_{m}^{2}}}\right)\frac{1}{\sqrt{2\pi\sigma_{x}^{2}}}e^{-\frac{\left(x-\mu_{x}\right)^{2}}{2\sigma_{x}^{2}}}dx\\
 & = & \frac{\lambda_{\text{max}}t}{\sigma_{x}}\sum_{m=1}^{M}\left(\frac{\sigma_{m}}{\alpha_{m}^{4}}e^{-\frac{\left(c_{m}-\mu_{x}\right)^{2}}{2\left(\alpha_{m}^{2}+\sigma_{x}^{2}\right)}}\int\limits _{-\infty}^{\infty}\left(x-c_{m}\right)^{2}\frac{e^{-\frac{\left(x-\mu_{m}\right)^{2}}{2\sigma_{m}^{2}}}}{\sqrt{2\pi\sigma_{m}^{2}}}dx\right),\end{eqnarray*}
where\[
\mu_{m}=\frac{\alpha_{m}^{2}\mu_{x}+\sigma_{x}^{2}c_{m}}{\alpha_{m}^{2}+\sigma_{x}^{2}}\qquad;\qquad\sigma_{m}^{2}=\frac{\alpha_{m}^{2}\sigma_{x}^{2}}{\alpha_{m}^{2}+\sigma_{x}^{2}}.\]
By defining $Y_{m}\sim\mathcal{N}(\mu_{m},\sigma_{m}^{2}),\quad(m=1,2,\ldots,M),$
the integral becomes $\mathbb{E}[Y_{m}^{2}-2c_{m}Y_{m}+c_{m}^{2}]=\sigma_{m}^{2}+\mu_{m}^{2}-2c_{m}\mu_{m}+c_{m}^{2}$,
and by substituting $\mu_{m},\sigma_{m}^{2}$ we get\[
\mathbb{E}\left[\mathcal{J}\left(X\right)\right]=\sum_{m=1}^{M}\frac{\lambda_{\text{max}}t\cdot e^{-\frac{\left(c_{m}-\mu_{x}\right)^{2}}{2\left(\alpha_{m}^{2}+\sigma_{x}^{2}\right)}}}{\alpha_{m}\left(\alpha_{m}^{2}+\sigma_{x}^{2}\right)^{\nicefrac{5}{2}}}\left[\sigma_{x}^{2}\left(\sigma_{x}^{2}+\alpha_{m}^{2}\right)+\alpha_{m}^{2}\left(c_{m}-\mu_{x}\right)^{2}\right].\]
\\

In the multi-dimensional case $\mathbf{X}\sim\mathcal{N}(\boldsymbol{\mu}_{x},\Sigma_{x})$,
the tuning functions are of the form\[
\lambda_{m}\left(\mathbf{x}\right)=\lambda_{\max}e^{-\frac{1}{2}\left(\mathbf{x}-\mathbf{c}_{m}\right)^{T}A_{m}^{-1}\left(\mathbf{x}-\mathbf{c}_{m}\right)},\qquad\left(m=1,\ldots,M\right),\]
and the population's Fisher information is given by\begin{eqnarray*}
\mathcal{J}\left(\mathbf{X}\right) & = & \mathbb{E}\left[\left(\sum_{m=1}^{M}\left(N_{t}^{m}-\lambda_{m}\left(\mathbf{X}\right)t\right)A_{m}^{-1}\left(\mathbf{X}-\mathbf{c}_{m}\right)\right)\left(\sum_{n=1}^{M}\left(N_{t}^{n}-\lambda_{n}\left(\mathbf{X}\right)t\right)A_{n}^{-1}\left(\mathbf{X}-\mathbf{c}_{n}\right)\right)^{T}\right]\\
 & = & \sum_{m=1}^{M}\sum_{n=1}^{M}\mathbb{E}\left[\left(N_{t}^{m}-\lambda_{m}\left(\mathbf{X}\right)t\right)\left(N_{t}^{n}-\lambda_{n}\left(\mathbf{X}\right)t\right)\middle|X\right]A_{m}^{-1}\left(\mathbf{X}-\mathbf{c}_{m}\right)\left(\mathbf{X}-\mathbf{c}_{n}\right)^{T}A_{n}^{-1}\\
 & = & \lambda_{\text{max}}t\sum_{m=1}^{M}A_{m}^{-1}\left\{ \left(\mathbf{X}-\mathbf{c}_{m}\right)\left(\mathbf{X}-\mathbf{c}_{m}\right)^{T}e^{-\frac{1}{2}\left(\mathbf{X}-\mathbf{c}_{m}\right)^{T}A_{m}^{-1}\left(\mathbf{X}-\mathbf{c}_{m}\right)}\right\} A_{m}^{-1}.\end{eqnarray*}
By virtue of the linearity, we calculate the expected value of $\mathcal{J}(\mathbf{X})$
with respect to $X$ by taking expectation over the expression in
curly brackets:\begin{eqnarray*}
\mathbb{E}\left\{ \cdot\right\}  & = & \int\left(\mathbf{X}-\mathbf{c}_{m}\right)\left(\mathbf{X}-\mathbf{c}_{m}\right)^{T}e^{-\frac{1}{2}\left(\mathbf{X}-\mathbf{c}_{m}\right)^{T}A_{m}^{-1}\left(\mathbf{X}-\mathbf{c}_{m}\right)}\frac{e^{-\frac{1}{2}\left(\mathbf{X}-\boldsymbol{\mu}_{x}\right)^{T}\Sigma_{x}^{-1}\left(\mathbf{X}-\boldsymbol{\mu}_{x}\right)}}{\left(2\pi\right)^{\nicefrac{D}{2}}\sqrt{\left|\Sigma_{x}\right|}}d\mathbf{X}\\
 & = & e^{-\frac{1}{2}\xi_{m}}\frac{\sqrt{\left|\Sigma_{m}\right|}}{\sqrt{\left|\Sigma_{x}\right|}}\int\left(\mathbf{X}-\mathbf{c}_{m}\right)\left(\mathbf{X}-\mathbf{c}_{m}\right)^{T}\frac{e^{-\frac{1}{2}\left(\mathbf{X}-\boldsymbol{\mu}_{m}\right)^{T}\Sigma_{m}^{-1}\left(\mathbf{X}-\boldsymbol{\mu}_{m}\right)}}{\left(2\pi\right)^{\nicefrac{D}{2}}\sqrt{\left|\Sigma_{m}\right|}}d\mathbf{X}\\
 & = & e^{-\frac{1}{2}\xi_{m}}\frac{\sqrt{\left|\Sigma_{m}\right|}}{\sqrt{\left|\Sigma_{x}\right|}}\left(\Sigma_{m}+\boldsymbol{\mu}_{m}\boldsymbol{\mu}_{m}^{T}-2\mathbf{c}_{m}\boldsymbol{\mu}_{m}^{T}+\mathbf{c}_{m}\mathbf{c}_{m}^{T}\right),\end{eqnarray*}
where\[
\Sigma_{m}=\left(A_{m}^{-1}+\Sigma_{x}^{-1}\right)^{-1}\qquad;\qquad\boldsymbol{\mu}_{m}=\Sigma_{m}\left(A_{m}^{-1}\mathbf{c}_{m}+\Sigma_{x}^{-1}\boldsymbol{\mu}_{x}\right);\]

\[
\xi_{m}=\mathbf{c}_{m}^{T}A_{m}^{-1}\mathbf{c}_{m}+\boldsymbol{\mu}_{x}^{T}\Sigma_{x}^{-1}\boldsymbol{\mu}_{x}-\boldsymbol{\mu}_{m}^{T}\Sigma_{m}^{-1}\boldsymbol{\mu}_{m}.\]
Using the identity $(A^{-1}+B^{-1})^{-1}=A(A+B)^{-1}B$ (\citet{searle1982mau})
and the symmetry of all matrices it is a simple mathematical exercise
to show that\[
\mathbb{E}\left[\mathcal{J}\left(\mathbf{X}\right)\right]=\lambda_{\text{max}}t\sum_{m=1}^{M}e^{-\frac{1}{2}\xi_{m}}\frac{\sqrt{\left|A_{m}\right|}A_{m}^{-1}}{\sqrt{\left|A_{m}+\Sigma_{x}\right|}}\left(A_{m}+\Sigma_{x}\right)^{-2}\left[\Sigma_{x}\left(A_{m}+\Sigma_{x}\right)+A_{m}\left(\mathbf{c}_{m}-\boldsymbol{\mu}_{x}\right)\left(\mathbf{c}_{m}-\boldsymbol{\mu}_{x}\right)^{T}\right]\]
and $\xi_{m}=(\mathbf{c}_{m}-\boldsymbol{\mu}_{x})^{T}(A_{m}+\Sigma_{x})^{-1}(\mathbf{c}_{m}-\boldsymbol{\mu}_{x})$.

\section*{S2. Uniform mean firing rate of population}

In this section we approximate the mean firing rate of the entire
population of sensory cells for the case of equally spaced tuning
functions with uniform width. First, consider the case $M=\infty$,
for which the population's mean firing rate is given by\[
\lambda\left(x,\alpha\right)=\lambda_{\max}\sum_{m=-\infty}^{\infty}e^{-\frac{1}{2}\left(\frac{x-m\Delta c}{\alpha}\right)^{2}}.\]
The function $\lambda(x,\alpha)$ is continuous with respect to $x$,
and therefore when the spacing goes to 0, $\Delta c\cdot\lambda(x,\alpha)$
converges to the Riemann integral\[
\lim_{\Delta c\to0}\Delta c\cdot\lambda(x,\alpha)=\lambda_{\max}\int\limits _{-\infty}^{\infty}e^{-\frac{1}{2}\left(\frac{x-c}{\alpha}\right)^{2}}dc=\lambda_{\max}\sqrt{2\pi}\alpha\]
which is independent of $x$. This means that for any $\varepsilon>0$
there exists $\delta>0$ such that if $\Delta c/\alpha<\delta$ then
the relative ripple (digression from uniformity) in the total firing
rate\[
\frac{\max\lambda(x,\alpha)-\min\lambda(x,\alpha)}{\min\lambda(x,\alpha)}\]
is smaller than $\varepsilon$.

Now consider the case of finite $M$,\[
\lambda\left(x,\alpha;M\right)=\lambda_{\max}\sum_{m=-\left\lceil \frac{M-1}{2}\right\rceil }^{\left\lceil \frac{M-1}{2}\right\rceil }e^{-\frac{1}{2}\left(\frac{x-m\Delta c}{\alpha}\right)^{2}}.\]
Since the limit at $M\to\infty$ exists, for every $x$ and $\varepsilon>0$
there exists $M_{\varepsilon}(x)>0$ such that for all $M>M_{\varepsilon}(x)$
the difference between the infinite sum to the finite sum is less
than $\varepsilon$, and therefore $\lambda(x,\alpha;M)$ can be well
approximated by\[
\sum_{m=-\infty}^{\infty}\lambda_{m}\left(x\right)\approx\frac{\lambda_{\max}}{\Delta c}\int\limits _{-\infty}^{\infty}e^{-\frac{1}{2}\left(\frac{x-c}{\alpha}\right)^{2}}dc=\frac{\lambda_{\max}\sqrt{2\pi}\alpha}{\Delta c}\triangleq\lambda\left(\alpha\right).\]

\begin{figure}
\begin{tabular}{ll}
A & B\tabularnewline
\includegraphics[width=85mm]{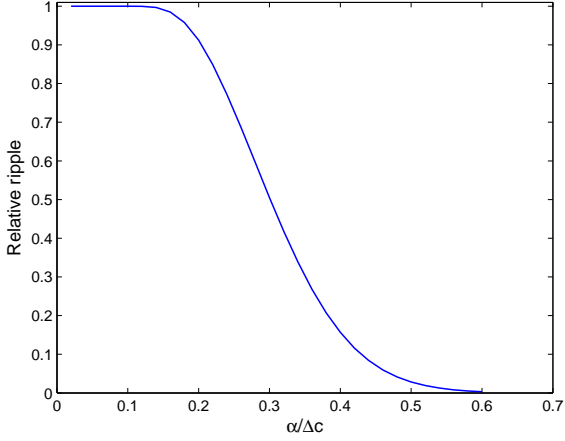} & \includegraphics[width=85mm]{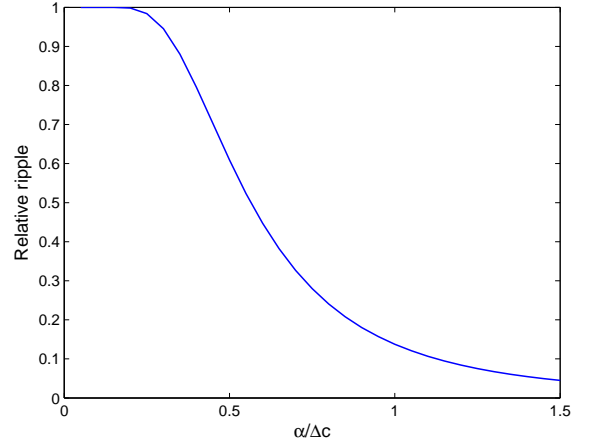}\tabularnewline
\end{tabular}\caption{\label{Flo:Ripple}The relative ripple in the mean firing rate of
a population of neurons with Gaussian tuning functions: (A) in single
dimension, (B) in two dimensions.}

\end{figure}

To quantify the desired ratio between the width and the spacing we
plot the relative ripple (figure \ref{Flo:Ripple}(A)). For $\alpha>0.583\Delta c$
the relative ripple falls down below 0.5\%, rendering the approximation
$\lambda(x,\alpha;M)\cong\lambda(\alpha)$ quite accurate. Note that
the approximation fails near the most extreme tuning functions, but
for a sufficiently large population this would not have any effect
on the results, because the probability of $X$ falling into these
regions would be vanishingly small for any rapidly-decaying prior.\\

The proof for the multi-dimensional case is similar, where the population's
mean firing rate is approximated by\[
\sum_{m=-\infty}^{\infty}\lambda_{m}\left(\mathbf{x}\right)\approx\frac{\lambda_{\max}}{\left(\Delta c\right)^{d}}\int\limits _{-\infty}^{\infty}e^{-\frac{1}{2}\left(\mathbf{x}-\mathbf{c}\right)^{T}A^{-1}\left(\mathbf{x}-\mathbf{c}\right)}d\mathbf{c}=\lambda_{\max}\left(\frac{\sqrt{2\pi}}{\Delta c}\right)^{d}\sqrt{|A|}\triangleq\lambda(A).\]
The conditions which make the approximation valid vary with the dimensionality
of the problem - as the dimension increases the width-spacing ratio
must be larger in order to maintain the same approximation accuracy.
For instance, in two dimensions where $A=\alpha\mathbf{I}$ (figure
\ref{Flo:Ripple}(B)), the relative ripple falls down below 0.5\%
only for $\alpha>1.446\Delta c$ .

\section*{S3. MMSE for multisensory integration}

In this section we derive the MMSE for the case of decoding bimodal
spike trains. Similarly to the unimodal noisy case, it is straightforward
to show that the joint posterior distribution of the environmental
state and noise variables is given by\begin{eqnarray*}
p\left(x,w_{v},w_{a}\middle|\mathbf{N}_{t}^{v},\mathbf{N}_{t}^{a}\right) & = & ce^{-\frac{\left(x-\mu_{x}\right)^{2}}{2\sigma_{x}^{2}}}e^{-\frac{w_{v}^{2}}{2\sigma_{w,v}^{2}}}e^{-\frac{w_{a}^{2}}{2\sigma_{w,a}^{2}}}e^{-\sum\limits _{m=1}^{M_{v}}\frac{\left(x+w_{v}-c_{m}^{v}\right)^{2}}{2\alpha_{v}^{2}}N_{t}^{m,v}}e^{-\sum\limits _{m=1}^{M_{a}}\frac{\left(x+w_{a}-c_{m}^{a}\right)^{2}}{2\alpha_{a}^{2}}N_{t}^{m,a}}\\
 & = & ce^{-\frac{\left(x-\mu_{x}\right)^{2}}{2\sigma_{x}^{2}}-\frac{w_{v}^{2}}{2\sigma_{w,v}^{2}}-\sum\limits _{m=1}^{M_{v}}\frac{\left(x+w_{v}-c_{m}^{v}\right)^{2}}{2\alpha_{v}^{2}}N_{t}^{m,v}}e^{-\frac{1}{2}\left(\frac{w_{a}^{2}}{\sigma_{w,a}^{2}}+\sum\limits _{m=1}^{M_{a}}\frac{\left(x+w_{a}-c_{m}^{a}\right)^{2}}{\alpha_{a}^{2}}N_{t}^{m,a}\right)},\end{eqnarray*}
which is Gaussian with respect to $W_{a}$, seeing that the argument
of the rightmost exponent is\[
-\frac{1}{2\hat{\sigma}_{w,a}^{2}}\left(w_{a}-\hat{\sigma}_{w,a}^{2}\sum\limits _{m=1}^{M_{a}}\frac{c_{m}^{a}-x}{\alpha_{a}^{2}}N_{t}^{m,a}\right)^{2}-\frac{1}{2}\left(\sum\limits _{m=1}^{M_{a}}\frac{\left(c_{m}^{a}-x\right)^{2}}{\alpha_{a}^{2}}N_{t}^{m,a}-\hat{\sigma}_{w,a}^{2}\left(\sum\limits _{m=1}^{M_{a}}\frac{c_{m}^{a}-x}{\alpha_{a}^{2}}N_{t}^{m,a}\right)^{2}\right),\]
where $\hat{\sigma}_{w,a}^{2}=(\sigma_{w,a}^{-2}+\alpha_{a}^{-2}\sum_{m}N_{t}^{m,a})^{-1}$.
Integration over $w_{a}$ gives the joint posterior distribution of
the state and the {}``visual'' noise:\begin{eqnarray*}
p\left(x,w_{v}\middle|\mathbf{N}_{t}^{v},\mathbf{N}_{t}^{a}\right) & = & ce^{-\frac{\left(x-\mu_{x}\right)^{2}}{2\sigma_{x}^{2}}-\frac{w_{v}^{2}}{2\sigma_{w,v}^{2}}-\sum\limits _{m=1}^{M_{v}}\frac{\left(x+w_{v}-c_{m}^{v}\right)^{2}}{2\alpha_{v}^{2}}N_{t}^{m,v}}e^{-\frac{1}{2}\left(\sum\limits _{m=1}^{M_{a}}\frac{\left(c_{m}^{a}-x\right)^{2}}{\alpha_{a}^{2}}N_{t}^{m,a}-\hat{\sigma}_{w,a}^{2}\left(\sum\limits _{m=1}^{M_{a}}\frac{\left(c_{m}^{a}-x\right)}{\alpha_{a}^{2}}N_{t}^{m,a}\right)^{2}\right)}\\
 & = & ce^{-\frac{\left(x-\mu_{x}\right)^{2}}{2\sigma_{x}^{2}}-\sum\limits _{m=1}^{M_{a}}\frac{\left(c_{m}^{a}-x\right)^{2}}{2\alpha_{a}^{2}}N_{t}^{m,a}+\hat{\sigma}_{w,a}^{2}\left(\sum\limits _{m=1}^{M_{a}}\frac{c_{m}^{a}-x}{2\alpha_{a}^{2}}N_{t}^{m,a}\right)^{2}}e^{-\frac{1}{2}\left(\frac{w_{v}^{2}}{\sigma_{w,v}^{2}}+\sum\limits _{m=1}^{M_{v}}\frac{\left(x+w_{v}-c_{m}^{v}\right)^{2}}{\alpha_{v}^{2}}N_{t}^{m,v}\right)},\end{eqnarray*}
which is Gaussian with respect to $W_{v}$, seeing that the argument
of the rightmost exponent is\[
-\frac{1}{2\hat{\sigma}_{w,v}^{2}}\left(w_{v}-\hat{\sigma}_{w,v}^{2}\sum_{m=1}^{M_{v}}\frac{c_{m}^{v}-x}{\alpha_{v}^{2}}N_{t}^{m,v}\right)^{2}-\frac{1}{2}\left(\sum\limits _{m=1}^{M_{v}}\frac{\left(c_{m}^{v}-x\right)^{2}}{\alpha_{v}^{2}}N_{t}^{m,v}-\hat{\sigma}_{w,v}^{2}\left(\sum_{m=1}^{M_{v}}\frac{c_{m}^{v}-x}{\alpha_{v}^{2}}N_{t}^{m,v}\right)^{2}\right),\]
where $\hat{\sigma}_{w,v}^{2}=(\sigma_{w,v}^{-2}+\alpha_{v}^{-2}\sum_{m}N_{t}^{m,v})^{-1}$.
Integration over $w_{v}$ gives the marginal distribution of the state:\[
p\left(x\middle|\mathbf{N}_{t}^{v},\mathbf{N}_{t}^{a}\right)=ce^{-\frac{\left(x-\mu_{x}\right)^{2}}{2\sigma_{x}^{2}}-\sum\limits _{m=1}^{M_{a}}\frac{\left(c_{m}^{a}-x\right)^{2}}{2\alpha_{a}^{2}}N_{t}^{m,a}+\hat{\sigma}_{w,a}^{2}\left(\sum\limits _{m=1}^{M_{a}}\frac{c_{m}^{a}-x}{2\alpha_{a}^{2}}N_{t}^{m,a}\right)^{2}-\sum\limits _{m=1}^{M_{v}}\frac{\left(c_{m}^{v}-x\right)^{2}}{2\alpha_{v}^{2}}N_{t}^{m,v}+\hat{\sigma}_{w,v}^{2}\left(\sum_{m=1}^{M_{v}}\frac{c_{m}^{v}-x}{2\alpha_{v}^{2}}N_{t}^{m,v}\right)^{2}},\]
which can be shown to be Gaussian with variance

\[
\hat{\sigma}^{2}=\left(\frac{1}{\sigma_{x}^{2}}+\frac{\sum_{m}N_{t}^{m,v}}{\alpha_{v}^{2}}+\frac{\sum_{m}N_{t}^{m,a}}{\alpha_{a}^{2}}-\frac{\hat{\sigma}_{w,v}^{2}\left(\sum_{m}N_{t}^{m,v}\right)^{2}}{\alpha_{v}^{4}}-\frac{\hat{\sigma}_{w,a}^{2}\left(\sum_{m}N_{t}^{m,a}\right)^{2}}{\alpha_{a}^{4}}\right)^{-1}.\]
By defining $Y_{v}\triangleq\sum_{m}N_{t}^{m,v}\sim\text{Pois}(\alpha_{v}t_{\text{eff}}^{v})$
and $Y_{a}\triangleq\sum_{m}N_{t}^{m,a}\sim\text{Pois}(\alpha_{a}t_{\text{eff}}^{a})$,
we get a relatively simple expression for the MMSE:\begin{eqnarray*}
\text{MMSE} & = & \mathbb{E}\left[\left(\frac{1}{\sigma_{x}^{2}}+\frac{Y_{v}}{\alpha_{v}^{2}}+\frac{Y_{a}}{\alpha_{a}^{2}}-\frac{\sigma_{w,v}^{2}Y_{v}^{2}}{\alpha_{v}^{2}\left(\alpha_{v}^{2}+\sigma_{w,v}^{2}Y_{v}\right)}-\frac{\sigma_{w,a}^{2}Y_{a}^{2}}{\alpha_{a}^{2}\left(\alpha_{a}^{2}+\sigma_{w,a}^{2}Y_{a}\right)}\right)^{-1}\right]\\
 & = & \mathbb{E}\left[\left(\frac{1}{\sigma_{x}^{2}}+\frac{Y_{v}}{\alpha_{v}^{2}+\sigma_{w,v}^{2}Y_{v}}+\frac{Y_{a}}{\alpha_{a}^{2}+\sigma_{w,a}^{2}Y_{a}}\right)^{-1}\right].\end{eqnarray*}

\section*{S4. Dynamics of the optimal width process}

In this section we prove that for the optimal piecewise-constant width
process defined in Section 2.3,\[
\underset{\Delta t\to0}{\lim}\alpha_{0}=\sigma_{x}\]
and\[
\underset{\Delta t\to0}{\lim}\frac{\alpha_{i}}{\alpha_{i+1}}=\sqrt{2},\quad i=0,1,\ldots,\]
where $\alpha_{i}$ is the value of the width process after obtaining
a total of $i$ spikes.\\

For short intervals, the Poisson random variable $Y_{\Delta t}=\sum_{m}N_{\Delta t}^{m}$
converges in probability to a Bernoulli random variable with success
probability $\alpha_{0}\Delta t_{\text{eff}}$. The MMSE at $t=\Delta t$
then takes the form\begin{eqnarray*}
\text{MMSE} & = & \sum_{k=0}^{1}\frac{1}{\sigma_{x}^{-2}+k\alpha_{0}^{-2}}P\left(Y_{\Delta t}=k\right)=\frac{1-\alpha_{0}\Delta t_{\text{eff}}}{\sigma_{x}^{-2}}+\frac{\alpha_{0}\Delta t_{\text{eff}}}{\sigma_{x}^{-2}+\alpha_{0}^{-2}}\\
 & = & \frac{\sigma_{x}^{-2}+\alpha_{0}^{-2}-\alpha_{0}^{-1}\Delta t_{\text{eff}}}{\sigma_{x}^{-2}+\alpha_{0}^{-2}}=1-\Delta t_{\text{eff}}\frac{\alpha_{0}^{-1}}{\alpha_{0}^{-2}+\sigma_{x}^{-2}}.\end{eqnarray*}
The MMSE is minimized when $\alpha_{0}^{-1}/(\alpha_{0}^{-2}+\sigma_{x}^{-2})$
is maximized, and thus $\alpha_{0}^{\text{opt}}=\sigma_{x}$. Given
the spikes history at $t=\Delta t$, if $Y_{\Delta t}=0$ then the
MMSE expression for the next interval remains unchanged, whereas if
$Y_{\Delta t}=1$ then the MMSE at the end of the next interval is
given by\[
\text{MMSE}=\mathbb{E}\left[\frac{1}{\sigma_{x}^{-2}+\left(\alpha_{0}^{\text{opt}}\right)^{-2}+\left(Y_{2\Delta t}-1\right)\alpha_{1}^{-2}}\right]=\mathbb{E}\left[\frac{1}{\sigma_{1}^{-2}+\left(Y_{2\Delta t}-1\right)\alpha_{1}^{-2}}\right],\]
where $\sigma_{1}=\sigma_{x}/\sqrt{2}$ is the effective uncertainty
after observing a single spike, and $Y_{2\Delta t}-1$ is a Bernoulli
random variable with success probability $\alpha_{1}\Delta t_{\text{eff}}$.
Applying the same analysis as before, $\alpha_{1}^{\text{opt}}=\sigma_{1}=\alpha_{0}^{\text{opt}}/\sqrt{2}$.
By recursion, one can see that as long as $k$ spikes were observed
the optimal width for the next interval remains $\alpha_{k}^{\text{opt}}$,
and if another spike is observed the optimal width for the next interval
becomes $\alpha_{k+1}^{\text{opt}}=\alpha_{k}^{\text{opt}}/\sqrt{2}.$

\pagebreak
\bibliographystyle{elsarticle-harv}
\bibliography{Yaeli_References}

\end{document}